\documentclass[onecolumn,showpacs,prc,superscriptaddress,amsmath,amssymb,preprintnumbers]{revtex4-1}

\newcommand{ \be }{\begin{eqnarray}}
\newcommand{ \ee }{\end{eqnarray}}
\newcommand{ \la }{\langle}
\newcommand{ \ra }{\rangle}

\newcommand{ \mean }[1]{\la #1 \ra}

\newcommand{ \Num }[2]{\mathrm{N}\mean{#1}_{#2}}
\newcommand{ \Den }[2]{\mathrm{D}\mean{#1}_{#2}}

\usepackage{dcolumn} 
\usepackage{bm} 

\usepackage{graphicx}
\usepackage{color}
\definecolor{dgreen}{cmyk}{1.,0.,1.,0.4}        
\definecolor{orange}{cmyk}{0.,0.353,1.,0.}    

\usepackage[pdftex,bookmarks]{hyperref}

\begin{document}


\title{Generic framework for anisotropic flow analyses with multi-particle azimuthal correlations}
\author{Ante Bilandzic} 
\affiliation{Niels Bohr Institute, Blegdamsvej 17, 2100 Copenhagen, Denmark}
\author{Christian Holm Christensen}
\affiliation{Niels Bohr Institute, Blegdamsvej 17, 2100 Copenhagen, Denmark}
\author{Kristjan Gulbrandsen}
\affiliation{Niels Bohr Institute, Blegdamsvej 17, 2100 Copenhagen, Denmark}
\author{Alexander Hansen} 
\affiliation{Niels Bohr Institute, Blegdamsvej 17, 2100 Copenhagen, Denmark}
\author{You Zhou}
\affiliation{Nikhef, Science Park 105, 1098 XG Amsterdam, 
The Netherlands}
\affiliation{Utrecht University, P.O. Box 80000, 3508 TA Utrecht, 
The Netherlands}
\date{\today}

\begin{abstract}
We present a new generic framework which enables exact and fast evaluation of all multi-particle azimuthal correlations. The framework can be readily used along with a correction framework for systematic biases in anisotropic flow analyses due to various detector inefficiencies. A new recursive algorithm has been developed for higher order correlators for the cases where their direct implementation is not feasible. We propose and discuss new azimuthal observables for anisotropic flow analyses which can be measured for the first time with our new framework. The effect of finite detector granularity on multi-particle correlations is quantified and discussed in detail. We point out the existence of a systematic bias in traditional differential flow analyses which stems solely from the applied selection criteria on particles used in the analyses, and is also present in the ideal case when only flow correlations are present. Finally, we extend the applicability of our generic framework to the case of  differential multi-particle correlations. 
\end{abstract}

\pacs{25.75.Ld, 25.75.Gz, 05.70.Fh}

\maketitle

\section{Introduction}
\label{s:Introduction}

In relativistic heavy-ion collisions the azimuthal anisotropy of the produced particles as a function of transverse momentum has emerged as the most renowned observable to study the collective properties of nuclear matter~\cite{Ollitrault:1992bk}. Due to the collision geometry in non-central heavy-ion collisions, the initial volume containing the interacting nuclear matter is anisotropic in coordinate space. Of particular interest is the scenario in which the produced nuclear matter managed to thermalize in this anisotropic volume, causing its initial anisotropy from the coordinate space to be transfered via mutual interactions into the resulting and observable anisotropy in momentum space. We refer to this phenomenon in this work as {\it collective anisotropic flow}, or just simply as {\it flow}. Clearly, collective anisotropic flow is a direct probe of the degree of thermalization of the produced matter, and correspondingly an indirect probe of its transport properties (e.g. viscosity).   

Whatever its underlying cause is, the resulting anisotropic distribution in momentum space can always be expanded into Fourier series~\cite{Voloshin:1994mz}:
\begin{equation}
f(\varphi) = \frac{1}{2\pi}\left[1+2\sum_{n=1}^{\infty}v_n\cos[n(\varphi-\Psi_n)]\right]\,.
\label{eq:Fourier}
\end{equation}
The first few coefficients (harmonics) in the above series have by now been thoroughly studied by experimentalists as well as theorists: The first coefficient $v_1$, 
is usually referred to as {\it directed flow}, the second coefficient, 
$v_2$, is referred to as {\it elliptic flow}, the third coefficient, 
$v_3$ is referred to as {\it triangular flow}, etc. $\Psi_n$ denotes the {\it symmetry plane} of the harmonic $v_n$ (in general different harmonics will have different symmetry planes). $\varphi$ denotes the azimuthal angles of the produced particles. For the case of an idealized initial geometry in heavy-ion collisions, all symmetry planes coincide and are equal to the {\it reaction plane} of the collision (a plane spanned by the impact parameter and the beam axis). Given the above Fourier series expansion, one can show, using just the orthogonality properties of trigonometric functions, that
\begin{equation}
v_n = \left<\cos[n(\varphi\!-\!\Psi_n)]\right>\,,
\label{eq:vn}
\end{equation}
where angular brackets denote an average over all particles in an event.    
Due to only mathematical steps involved in its derivation, we stress that Eq.~(\ref{eq:vn}) {\it per se} has no physical meaning. In particular, Eq.~(\ref{eq:vn}) can give rise to non-vanishing flow harmonics $v_n$ irrespectively of whether the azimuthal anisotropy in the momentum distribution has its origin in collective anisotropic flow or in some other completely unrelated physical process which can also yield event-by-event anisotropies (e.g. mini-jets). We now attempt to attach a more rigorous treatment to the concept of ``collectivity" by discussing which tools and observables we can  utilize experimentally in order to disentangle it from processes which generally involve only a small subset of the produced particles, generally termed ``nonflow".  

In order to make a statement on whether the harmonics $v_n$ in Eq.~(\ref{eq:Fourier}) are dominated by contributions from collective anisotropic flow or by some other processes which are non-collective in nature, we can use correlation techniques involving two or more particles. In this paper our main focus will be on the latter, to which we refer to as {\it multi-particle correlation techniques}. When only collective anisotropic flow is present, all produced particles are independently emitted, and are correlated only to some common reference planes. This physical observation translates into the following mathematical statement:
\begin{equation}
f(\varphi_1,\ldots,\varphi_n) = f_{\varphi_{1}}(\varphi_{1})\cdots f_{\varphi_{n}}(\varphi_n)\,.
\label{eq:factorization}
\end{equation}
The left-hand side of Eq.~(\ref{eq:factorization}) is a joint multi-variate probability density function (p.d.f.) of $n$ observables $\varphi_1,\ldots,\varphi_n$. The right-hand side of Eq.~(\ref{eq:factorization}) is the product of the normalized marginalized p.d.f, $f_{\phi_i}(\phi_i)$, where $1\leq i\leq n$, which are the same~\cite{Danielewicz:1983we} and are given by Eq.~(\ref{eq:Fourier}). Therefore, when all particles are emitted independently, as is the case for collective anisotropic flow, the joint p.d.f. for {\it any} number of particles will factorize as in Eq.~(\ref{eq:factorization}). Based on this reasoning, one can build up, in principle, infinitely many independent azimuthal observables sensitive to various combinations of flow harmonic moments and corresponding symmetry planes by adding more and more particles to the observables. When flow harmonics fluctuate event-by-event, different underlying p.d.f.'s of flow fluctuations will result in different values of flow harmonic moments and corresponding symmetry planes. This illustrates our main point: In order to determine the underlying p.d.f. of flow fluctuations, one is necessarily led towards multi-particle correlation techniques. We will elaborate on this point in detail and generalize it further in the main part of the paper. For completeness, we now present the historical overview of the utilization of multi-particle correlation techniques in anisotropic flow analyses, together with all of the technical limitations and issues inherit to them, which this paper overcomes. 

Multi-particle correlation techniques in anisotropic flow analyses have been used for more than
three decades. In the theoretical studies of global event shapes~\cite{Danielewicz:1983we} and in the subsequent 
study presented in~\cite{Ollitrault:1992bk}, the joint multi-variate p.d.f. of $M$ particles for an event with multiplicity 
$M$ was utilized in flow analyses for the first time. On the other hand, the very first experimental attempt to go beyond two-particle azimuthal correlations~\cite{Wang:1991qh} date back to Bevalac work published in~\cite{Jiang:1992bw}. In that paper, a quantitative description of collectivity was attempted by generalizing the observable for two-particle correlations, namely the smaller angle between the transverse momenta of two produced particles, into the geometric mean of $n\ (n>2)$ azimuthal separations within the $n$-particle multiplet. However, it was realized immediately that the net contribution of low-order few-particle correlations is cumulative if one increases the number of particles in such multiplets, which triggered the demand for more sophisticated techniques that would instead suppress systematically such contributions for increasingly large multiplets~\cite{Jiang:1992bw}. 

This was pursued further in a series of papers on multi-particle correlations and cumulants by Borghini {\it et al} (for a summary of the mathematical and statistical properties of cumulants we refer the reader to~\cite{Kubo:1962}). In the first paper of the series~\cite{Borghini:2000sa}, Borghini {\it et al} defined cumulants in the context of flow analyses in terms of the moments of the distribution of the $Q$-vector amplitude~\cite{Ollitrault:1992bk,Barrette:1994xr,Voloshin:1994mz}. As a landmark of their approach, the authors have introduced a formalism of generating functions accompanied with interpolation methods in the complex plane as the simplest and fastest way to calculate cumulants from experimental data. The formalism of generating functions is particularly robust against biases stemming from non-uniform detector acceptance, which is frequently the dominant systematic bias in anisotropic flow analyses. However, there were some serious drawbacks, which were recognized and discussed already by the authors in the original paper. Most notably, both two- and multi-particle cumulants were plagued by trivial and non-negligible contributions from autocorrelations, which caused an interference between the various harmonics. This led the authors to propose an improved version of the generating function in~\cite{Borghini:2001vi}, which by design generated cumulants free from autocorrelations. In essence, the way cumulants were defined conceptually has changed between the two papers: In~\cite{Borghini:2001vi} cumulants were defined directly in terms of multi-particle azimuthal correlations, which are free from autocorrelations by definition, while in~\cite{Borghini:2000sa}  
cumulants were defined in terms of the moments of the distribution of the $Q$-vector amplitude, which by definition have contributions from autocorrelations. Both methods to calculate cumulants were capable of estimating reference and differential flow. Further improvement, still relying on the formalism of generating functions, came with the Lee-Yang zero (LYZ) method~\cite{Bhalerao:2003xf,Bhalerao:2003yq}, which isolates the genuine multi-particle estimate for flow harmonics, corresponding to the asymptotic behavior of the cumulant series. The formalism of generating functions, however, has its own built-in systematic biases. Most importantly, the proposed interpolating methods in the complex plane to calculate cumulants are not numerically stable for all values of flow harmonics and multiplicity (``parameter $r_0$ has to be tuned"); in addition, one never exactly recovers the cumulants as they are defined (``the series expansion of the generating functions has to be terminated manually at a certain order, in order to close the coupled system of equations for the cumulants"); finally, the formalism as presented in these papers is limited to the cases where all harmonics in multi-particle correlators coincide. A notable alternative cumulant approach in terms of implementation was used in~\cite{Adler:2002pu}, which, at the expense of reducing statistics, removed autocorrelations by explicitly constructing  multiple subevents from the original event.

These limitations were removed partially with $Q$-cumulants (QC) published recently in~\cite{Bilandzic:2010jr}, which do not rely on the formalism of generating functions, but instead utilize Voloshin's original idea of expressing  multi-particle azimuthal correlations analytically in terms of $Q$-vectors evaluated (in general) in different harmonics. $Q$-cumulants, however, are very tedious to calculate analytically, and such calculations were accomplished only for a rather limited subset of multi-particle azimuthal correlations which have been most frequently used in anisotropic flow analyses to date.

The present paper surpasses completely all technical limitations of these previous publications and provides a {\it generic framework} allowing {\it all} multi-particle azimuthal correlations to be evaluated analytically, with a fast single pass over the data, free from autocorrelations by definition, and corrected for systematic biases due to various detector inefficiencies (e.g. non-uniform azimuthal acceptance, $p_{\rm T}$-dependent reconstruction efficiency, finite detector granularity, etc.). With this framework, a plethora of new multi-particle azimuthal observables are now accessible experimentally. In this paper we propose and discuss some new concrete examples (so-called {\it standard candles}). We have paid special attention to the development of algorithms, which can be used to calculate recursively higher-order multi-particle azimuthal correlators in terms of lower-order ones, for the cases when their standalone generic formulae are too long and impractical for direct use and implementation. Finally, we point out the existence of a peculiar systematic bias in traditional differential flow analyses, when all particles are divided into the two groups of reference particles (RP) and particles of interest (POI). This systematic bias stems solely from the selection criteria for RPs and POIs, and is present also in the ideal case when all nonflow correlations are absent.    
    
The paper is organized as follows. In Section~\ref{s:Two- and multi-particle azimuthal correlations}, we introduce two- and multi-particle azimuthal correlations, motivate and discuss their usage in anisotropic flow analyses, and point out the technical issues which plagued their evaluation in the past. In Section~\ref{s:Generic equations}, we outline our new generic framework which enables exact and fast evaluation of all multi-particle azimuthal correlations, and can also be used to correct for systematic biases due to various detector inefficiencies. In Section~\ref{s:Monte Carlo studies} we use two toy Monte Carlo studies to demonstrate the framework's ability to correct for biases due to non-uniform azimuthal acceptance and non-uniform reconstruction efficiency. We then use a realistic Monte Carlo to demonstrate its usage in the measurement of some new flow observables that we propose and discuss in detail. In Section~\ref{s:Detectors with finite granularity}, we point out how biases due to finite granularity of the detector must be considered and corrected for in the measurement of  multi-particle azimuthal correlations. Finally, in Section~\ref{s:Systematic bias due to particle selection criteria}, we discuss the systematic bias which is present in traditional differential flow analyses even when all nonflow correlations are absent, but arise from the selection criteria of particles used for the differential flow analysis. In the  Appendices we present all technical steps in detail.  


\section{Two- and multi-particle azimuthal correlations}
\label{s:Two- and multi-particle azimuthal correlations}

We consider two- and multi-particle azimuthal correlations measured event-by-event as our basic observables whose moments can be related to moments of the flow harmonics and the corresponding symmetry planes. This relation can be illustrated with the simple example of the two-particle azimuthal correlation of harmonics $n$ and $-n$. For the dataset consisting of $M$ azimuthal angles $\varphi_1,\varphi_2,\ldots,\varphi_M$ we have:
\begin{eqnarray}
\left<2\right>_{n,-n} &\equiv& \left<e^{in(\varphi_1-\varphi_2)}\right> = \left<\cos n(\varphi_1-\varphi_2)\right>\nonumber\\
 &=& \frac{1}{M(M\!-\!1)}\,\sum_{\begin{subarray}{c}i,j=1\\i\neq j\end{subarray}}^{M}\cos n(\phi_i-\phi_j)\,.
\label{eq:2pBasicExample}
\end{eqnarray}
The constraint $i\neq j$ removes contributions from autocorrelations in each sum by definition. Using the factorization property in Eq.~(\ref{eq:factorization}) for the case of joint two-particle p.d.f. and using the orthogonality properties of trigonometric functions, one can show that the first and second moment of $\left<2\right>_{n,-n}$ are given as:
\begin{eqnarray}
\mu_{\left<2\right>_{n,-n}} &=& v_n^2\,,\label{eq:mu2p}\\
\sigma_{\left<2\right>_{n,-n}}^2 &=& \frac{1+v_{2n}^2}{M(M-1)}+2\frac{M-2}{M(M-1)}v_n^2(1+v_{2n})\nonumber\\
&&{}+\frac{(M-2)(M-3)}{M(M-1)}v_n^4-v_n^4\,.\label{eq:2pNoise}
\end{eqnarray}
These are the analytic expressions for the mean and variance of the two-particle azimuthal correlations, which are valid for the general case when the Fourier-like p.d.f. (\ref{eq:Fourier}) is parametrized with all harmonics $v_n$. 

Motivated with the previous simple example, we now introduce our main observables, namely multi-particle azimuthal correlations, in a generic way. The average $m$-particle correlation in harmonics $n_1,n_2,\ldots,n_m$ is given by the following generic definition:
\begin{eqnarray}
\left<m\right>_{n_1,n_2,\ldots,n_m}&\equiv&\left<e^{i(n_1\varphi_{k_1}+n_2\varphi_{k_2}+\cdots+n_m\varphi_{k_m})}\right>\nonumber\\
&\equiv&\frac{\displaystyle\sum_{\begin{subarray}{c}k_1,k_2,\ldots,k_m=1\\k_1\neq k_2\neq \ldots\neq k_m\end{subarray}}^{M}w_{k_1}w_{k_2}\cdots w_{k_m}\,e^{i(n_1\varphi_{k_1}+n_2\varphi_{k_2}+\cdots+n_m\varphi_{k_m})}}{\displaystyle\sum_{\begin{subarray}{c}k_1,k_2,\ldots,k_m=1\\k_1\neq k_2\neq \ldots\neq k_m\end{subarray}}^{M} w_{k_1}w_{k_2}\cdots w_{k_m}}\,.
\label{eq:mpCorrelation}
\end{eqnarray}
In the above definition, $M$ is the multiplicity of an event, $\varphi$ labels the azimuthal angles of the produced particles, while $w$ labels particle weights whose physical meaning and use cases will be elaborated on. We have in summation enforced the condition $k_1\neq k_2\neq \ldots\neq k_m$ in order to remove the trivial and non-negligible contributions from all possible autocorrelations (self-correlations) by definition in all summands. We stress that we consider any correlation technique utilized in anisotropic flow analyses to be unsound and unusable if it has any kind of contribution stemming from autocorrelations.  

Particle weights appearing in definition (\ref{eq:mpCorrelation}) can be used to remove systematic biases originating from detector inefficiencies of various types. Well known examples of particle weights are so-called $\varphi$-weights, $w_{\varphi}$, which deal with the systematic bias due to non-uniform acceptance in azimuth, and $p_{\rm T}$-weights, $w_{p_{\rm T}}$, which deal with the non-uniform transverse momentum reconstruction efficiency of produced particles. In general, we allow the particle weight $w$ to be the most general function of the azimuthal angle, transverse momentum, pseudorapidity, particle type, etc.:
\begin{equation}
w = w(\varphi,p_{{\rm T}},\eta,{\rm PID},\ldots)\,.
\label{eq:generalParticleWeights}
\end{equation}
The new generic framework presented in this paper allows one to use the above general particle weights for any multi-particle azimuthal correlation. In subsequent sections in toy Monte Carlo studies we provide two concrete examples. 

We can straightforwardly relate various moments of the observables defined in Eq.~(\ref{eq:mpCorrelation}) to various moments of the harmonics $v_n$ and the  symmetry planes $\Psi_n$. In particular, relying solely on factorization as in Eq.~(\ref{eq:factorization}) and orthogonality properties of trigonometric functions, the following analytic expression follows for the first moment:
\begin{equation}
\mu_{\left<m\right>_{n_1,n_2,\ldots,n_m}} \equiv 
\left<e^{i(n_1\varphi_1+\cdots+n_m\varphi_{m})}\right> = v_{n_1}\cdots v_{n_m}e^{i(n_1\Psi_{n_{1}}+\cdots+n_m\Psi_{n_m})}\,.
\label{eq:mixedHarmonicsExpVersion}
\end{equation}
This result was first presented in~\cite{Bhalerao:2011yg}. When the averaging is extended to all events, only the isotropic correlators, i.e. the ones for which $n_1+n_2+\cdots+n_m = 0$, will have non-zero values~\cite{Bhalerao:2011yg}. It is obvious from the expression~(\ref{eq:mixedHarmonicsExpVersion}) that the trivial periodicity of each symmetry plane is automatically accounted for. As already remarked in the introduction, for the case of an idealized initial geometry all symmetry planes $\Psi_n$ coincide and the imaginary part of Eq.~(\ref{eq:mixedHarmonicsExpVersion}) is identically zero for isotropic correlators. However, we point out that, in the more realistic case, the effects of flow fluctuations can be independently quantified by measuring the imaginary parts of isotropic correlators in mixed harmonics as well, which a priori are non-vanishing. The importance of our new generic framework is that it makes it possible for the first time to measure the above observables~(\ref{eq:mixedHarmonicsExpVersion}) for any number of particles $m$ in the correlators, for any values of the harmonics $n_1, n_2,\ldots,n_m$, and for both the real and imaginary parts. 

One of the consequences of event-by-event flow fluctuations is the fact that $\left<v_n^k\right>\neq \left<v_n\right>^k$,
where flow moments $\left<v_n^k\right>$ are defined as
\begin{equation}
\left<v_n^k\right> \equiv \int v_n^k\ f(v_n)\ dv_n\,.
\end{equation}
Different underlying p.d.f.'s, $f(v_n)$, of event-by-event flow fluctuations will yield different values for the moments $\left<v_n^k\right>$. Looking at this statement from a different angle, we can also conclude that two completely different p.d.f.'s, reflecting completely different physical mechanisms that drive flow fluctuations, can have, accidentally, the very same first moment $\left<v_n\right>$. Thus, the traditional way of reporting results of anisotropic flow analyses by estimating only the first moment of the underlying p.d.f, namely $\left<v_n\right>$, is, from our point of view, rather incomplete. Instead, one should measure as many moments $\left<v_n^k\right>$ as possible of the underlying p.d.f, $f(v_n)$, because each moment by construction carries independent information. To finalize this discussion, we stress that a priori it is not guaranteed that a p.d.f. is uniquely determined by its moments. Necessary and sufficient conditions for the p.d.f. to be uniquely determined in terms of its moments have been worked out only recently and are known as the Krein-Lin conditions~\cite{Stoyanov:2006}:
\begin{equation}
K[f]\equiv\int_0^\infty\frac{-\ln f(x^2)}{1+x^2}\, dx\quad\Rightarrow\quad K[f]=\infty\,,
\label{eq:Krein}
\end{equation}
\begin{equation}
L(x)\equiv-\frac{xf'(x)}{f(x)}\quad\Rightarrow\quad \lim_{x\rightarrow\infty}L(x)=\infty\,.
\label{eq:Lin}
\end{equation}
The generic framework presented in this paper enables one to measure the flow moments $\left<v_n^k\right>$ for any $k$. Such results, in combination with the Krein-Lin conditions outlined above, can be used to experimentally constrain the nature of the p.d.f. for flow fluctuations. 


\section{Generic equations}
\label{s:Generic equations}

In this section, we present and discuss our main results. For an event with multiplicity $M$ we construct the following two sets:
\begin{eqnarray}
\mathrm{azimuthal\ angles}&:& \{\varphi_1,\varphi_2,\ldots,\varphi_M\}\,,\nonumber\\
\mathrm{weights} &:& \{w_1,w_2,\ldots,w_M\}\,,
\label{eq:sets}
\end{eqnarray}
where $\varphi$ labels the azimuthal angles of particles, while $w$ labels particle weights introduced in Eq.~(\ref{eq:generalParticleWeights}). Given these two sets, we calculate in each event weighted $Q$-vectors~\cite{Ollitrault:1992bk,Barrette:1994xr,Voloshin:1994mz} as complex numbers defined by
\begin{equation}
Q_{n,p} \equiv \sum_{k=1}^{M}w_k^p\,e^{in\varphi_k} \,.
\label{eq:Qvector}
\end{equation}
From the above definition, it immediately follows that:
\begin{equation}
Q_{-n,p} = Q_{n,p}^*\,,
\end{equation}
which shall be used in the implementation of our final results in order to reduce the amount of needed computations. We remark that we need a single pass over the particles to calculate the $Q$-vectors for multiple values of indices $n$ and $p$.  

We first observe that the expressions in the numerator
and the denominator of Eq.~(\ref{eq:mpCorrelation}) are trivially related. Namely, given the result for the numerator which depends on harmonics $n_1,n_2,\ldots,n_m$, the result for the denominator can be obtained by using the result for numerator and setting all harmonics $n_1,n_2,\ldots,n_m$ to 0. Therefore in what follows we focus mostly on the results for the numerator, and introduce the following shortcuts: 
\begin{eqnarray}
\Num{m}{n_1,n_2,\ldots,n_m}&\equiv&\displaystyle\sum_{\begin{subarray}{c}k_1,k_2,\ldots,k_m=1\\k_1\neq k_2\neq \ldots\neq k_m\end{subarray}}^{M}\!\!\!\!\!w_{k_1}w_{k_2}\cdots w_{k_m}\,e^{i(n_1\varphi_{k_1}+n_2\varphi_{k_2}+\cdots+n_m\varphi_{k_m})}\,,\label{eq:num}\\
\Den{m}{n_1,n_2,\ldots,n_m}&\equiv&\displaystyle\sum_{\begin{subarray}{c}k_1,k_2,\ldots,k_m=1\\k_1\neq k_2\neq \ldots\neq k_m\end{subarray}}^{M}\!\!\!\!\!w_{k_1}w_{k_2}\cdots w_{k_m}\label{eq:den}\\
&=&\Num{m}{0,0,\ldots,0}\,.\label{eq:denNum0000}
\end{eqnarray}
%

The key experimental question in anisotropic flow analyses relying on correlation techniques was how to enforce the condition $k_1\neq k_2\neq \ldots\neq k_m$ in the summations (\ref{eq:num}) and (\ref{eq:den}) without using the brute force approach of $m$ nested loops. Such an approach is not feasible even for four-particle correlators and events with a multiplicity of the order of 100 particles. It is therefore unusable for events with multiplicities of the order of 1000 particles, characteristic of present day relativistic heavy-ion collisions. How this problem was resolved approximately and for some specific correlators has been summarized in Section~\ref{s:Introduction}. Here we provide an exact and general answer. 

We outline explicitly the results for the case of 2-, 3-, and 4-p correlators expressed analytically in terms of $Q$-vectors defined in Eq.~(\ref{eq:Qvector}). For 2-p correlators it follows:
\begin{eqnarray}
\Num{2}{n_1,n_2}&=&Q_{n_1,1} Q_{n_2,1}-Q_{n_1+n_2,2}\,,\nonumber\\
\Den{2}{n_1,n_2}&=&\Num{2}{0,0}\nonumber\\ 
&=&Q_{0,1}^2-Q_{0,2}\,. 
\label{eq:2pCorrelation}
\end{eqnarray}
Additionally, for 3-p correlators it follows:
\begin{eqnarray}
\Num{3}{n_1,n_2,n_3}&=&Q_{n_1,1} Q_{n_2,1} Q_{n_3,1}-Q_{n_1+n_2,2} Q_{n_3,1}-Q_{n_2,1} Q_{n_1+n_3,2}\nonumber\\
&&{}-Q_{n_1,1} Q_{n_2+n_3,2}+2 Q_{n_1+n_2+n_3,3}\,,\nonumber\\
\Den{3}{n_1,n_2,n_3}&=&\Num{3}{0,0,0}\nonumber\\
&=&Q_{0,1}^3-3Q_{0,2} Q_{0,1}+2 Q_{0,3}\,.
\label{eq:3pCorrelation}
\end{eqnarray}
Finally, for 4-p correlators we have obtained:
\begin{eqnarray}
\Num{4}{n_1,n_2,n_3,n_4}&=&Q_{n_1,1} Q_{n_2,1} Q_{n_3,1} Q_{n_4,1}-Q_{n_1+n_2,2} Q_{n_3,1} Q_{n_4,1}
-Q_{n_2,1} Q_{n_1+n_3,2} Q_{n_4,1}\nonumber\\
&&{}-Q_{n_1,1} Q_{n_2+n_3,2} Q_{n_4,1}+2 Q_{n_1+n_2+n_3,3} Q_{n_4,1}-Q_{n_2,1}
Q_{n_3,1} Q_{n_1+n_4,2}\nonumber\\
&&{}+Q_{n_2+n_3,2} Q_{n_1+n_4,2}
-Q_{n_1,1} Q_{n_3,1} Q_{n_2+n_4,2}+Q_{n_1+n_3,2} Q_{n_2+n_4,2}\nonumber\\
&&{}+2 Q_{n_3,1} Q_{n_1+n_2+n_4,3}
-Q_{n_1,1} Q_{n_2,1} Q_{n_3+n_4,2}+Q_{n_1+n_2,2}
Q_{n_3+n_4,2}\nonumber\\
&&{}+2 Q_{n_2,1} Q_{n_1+n_3+n_4,3}+2 Q_{n_1,1} Q_{n_2+n_3+n_4,3}-6 Q_{n_1+n_2+n_3+n_4,4}\,,\\
\Den{4}{n_1,n_2,n_3,n_4}&=&\Num{4}{0,0,0,0}\nonumber\\
&=&
Q_{0, 1}^4 - 6 Q_{0, 1}^2 Q_{0, 2} + 3 Q_{0, 2}^2 + 8 Q_{0, 1} Q_{0, 3} - 
 6 Q_{0, 4}\,.
\label{eq:4pCorrelation} 
\end{eqnarray}
The analogous results for higher order correlators can be spelled out in a similar manner, but they are too long to fit in this paper. Instead, we provide them calculated and implemented (in .cpp and .nb file formats) up to and including 8-particle correlators at the following link~\cite{CHC}. As an alternative, we have developed recursive algorithms which, at the expense of runtime performance, calculate analytically higher order correlators in terms of lower order ones. The recursive algorithms will be presented in detail in Section~\ref{ss:Algorithm}. 

As the number of particles in correlators increases, the above analytical standalone expressions for multi-particle correlators quickly become  impractical for direct use and implementation. For instance, the analogous analytic result for the 8-p correlator contains 4140 distinct terms, each of which is a product of up to eight distinct complex $Q$-vectors. A closer look at the structure of these analytic solutions revealed that the number of distinct terms per correlator form a well known Bell sequence:
\begin{equation}
 1, 2, 5, 15, 52, 203, 877, 4140, 21147, \ldots ,
\label{eq:bell_sequence}
\end{equation}
which gives the number of different ways to partition a set with $m$ elements. In our context, $m$ is the number of particles in the correlator, and ``different way to partition" corresponds to different possible contributions from autocorrelations.  

The above results can be straightforwardly extended to the case of differential multi-particle correlators, for which one particle in the multiplet is restricted to belong only to the narrow differential bin of interest; the self-contained treatment of  
differential multi-particle correlators is presented in Appendix~\ref{s:Appendix to differential multi-particle correlators}.


\subsection{Algorithm}
\label{ss:Algorithm}

As already remarked, direct evaluation of expression~\eqref{eq:num} for higher order correlators quickly becomes impractical
due to the number of terms.  For that reason, we have developed
algorithms which recursively express all higher order correlators in
terms of the lower order ones. Observing that
\begin{align*}
  \Num{1}{n_1} &= Q_{n_1,1}\,,\\
  \Num{2}{n_1,n_2} &= \Num{1}{n_1} Q_{n_2,1} - Q_{n_1+n_2,2}\,,\\
  \Num{3}{n_1,n_2,n_3} &= \Num{2}{n_1,n_2} Q_{n_3,1}
  - \Num{1}{n_1} Q_{n_2+n_3,2} - \Num{1}{n_2} Q_{n_1+n_3,2}
  + 2 Q_{n_1+n_2+n_3,3}\,,\\
\Num{4}{n_1,n_2,n_3,n_4} &= \Num{3}{n_1,n_2,n_3} Q_{n_4,1}
- \Num{2}{n_1,n_2} Q_{n_3+n_4,2}
- \Num{2}{n_1,n_3} Q_{n_2+n_4,2}
- \Num{2}{n_2,n_3} Q_{n_1+n_4,2}\nonumber\\
&\quad{} + 2 \Num{1}{n_1} Q_{n_2+n_3+n_4,3}
+ 2 \Num{1}{n_2} Q_{n_1+n_3+n_4,3}
+ 2 \Num{1}{n_3} Q_{n_1+n_2+n_4,3}
- 6 Q_{n_1+n_2+n_3+n_4,4}\,,
\end{align*}
it is clear that the $\Num{m}{n_1,\ldots,n_m}$ is determined through
ordered partitions of the numbers $\{n_1,\ldots,n_m\}$.  We can use
this property to calculate $\Num{m}{n_1,\ldots,n_m}$ for \emph{any}
$m$ as outlined in pseudo--code in~\eqref{eq:algo1}
\begin{tabbing}
  \hspace*{2em}\=\hspace*{2em}\=\hspace*{2em}\kill
  \>$\Num{1}{n_1}^\prime$: return $Q_{n_1,1}$\\
  \>$\Num{m}{n_1,\ldots,n_m}^{\prime}$:\\
  \>\>$C \leftarrow 0$\\
  \>\>for \= $k\leftarrow(m-1),1$ do\\
  \>\>\>for \= each combination $c=\{c_1,\ldots,c_k\}$ of $\{n_1,\ldots,n_{m-1}\}$ do\\
  \>\>\>\>$q \leftarrow \sum_{j\,\text{not in}\,c}n_j$\\
  \>\>\>\>$C\leftarrow C +  (-1)^{m-k}\,(m-k-1)!\times\Num{k}{c_1,\ldots,c_k}^{\prime}\times Q_{q,m-k}$\\
  \>\>\> end for each $c$\\
  \>\>end for $k$\\
  \>\>return $C\quad.$
  \refstepcounter{equation}\label{eq:algo1}\>\>\`(\arabic{equation})
\end{tabbing}
A different recursive relation can be developed by examining Eq.~(\ref{eq:num}) itself. It can be seen that the innermost sum can be rewritten without the constraint of not being equal to any other index in the following way: 
\begin{eqnarray}
\Num{m}{n_1,n_2,\ldots,n_m}&=&\sum_{\begin{subarray}{c}k_1,k_2,\ldots,k_{m-1}=1\\k_1\neq k_2\neq \ldots\neq k_{m-1}\end{subarray}}^{M}w_{k_1}w_{k_2}\cdots w_{k_{m-1}}\,e^{i(n_1\varphi_{k_1}+n_2\varphi_{k_2}+\cdots+n_{m-1}\varphi_{k_{m-1}})}\nonumber\\
&&\ \ \ \ \ \ \ \ \ \ \ \ \ \times \bigg( \sum_{k_m = 1}^M w_{k_m}e^{in_m \varphi_{k_m}}-\sum_{j = 1}^{m-1}w_{k_j} e^{in_m\varphi_{k_j}}  \bigg)\,.
\label{eq:mpCorrelationRecursion_1}
\end{eqnarray}
This can be expanded into the following recursive formula, where, however, one must be careful to set the power of the weights equal to the number of summands (i.e. $n_i+n_j$ would have a corresponding $w^2$ term, $n_i+n_j+n_k$ would have a corresponding $w^3$ term, etc.):
\begin{eqnarray}
\Num{m}{n_1,n_2,\ldots,n_m}&=&Q_{n_m,1}\Num{m-1}{n_1,n_2,\ldots,n_{m-1}}-\Num{m-1}{n_1+n_m,n_2,\ldots,n_{m-1}}\nonumber\\
&&{}-\Num{m-1}{n_1,n_2+n_m,\ldots,n_{m-1}}-\ldots-\Num{m-1}{n_1,n_2,\ldots,n_{m-1}+n_m}\,.
\label{eq:mpCorrelationRecursion_2}
\end{eqnarray}
An optimized version of this recursive formula, which ensures that unique terms are evaluated only once, is shown in pseudo--code in~\eqref{eq:algo2}, where initially all $c_i=1$
  \begin{tabbing}
    \hspace*{2em}\=$\Num{1}{n_1}^{\prime\prime}(\{c_1\})$: return $Q_{n_1,c_1}$\\
    \>$\Num{m}{n_1,\ldots,n_m}^{\prime\prime}(\{c_1,\ldots,c_m\})$:\\
    \>\hspace*{2em}\=$C \leftarrow Q_{n_m,c_m} \times
    \Num{m-1}{n_1,\ldots,n_{m-1}}^{\prime\prime}(\{c_1,\ldots,c_{m-1}\})$\\
    \>\>if \= $c_{m} \leq 1$ then\\
    \>\>\>for \=$i\leftarrow 1,m-1$ do\\
    \>\>\>\>$C\leftarrow C  
    -c_i\times\Num{m-1}{n_1,\ldots,n_{i}+n_{m},\ldots , n_{m-1}}^{\prime\prime}(\{c_1,\ldots,c_i+1,\ldots,c_{m-1}\})$\\ 
    \>\>\>end for $i$\\
    \>\>end if\\
    \>\>return $C\quad.$
    \refstepcounter{equation}\label{eq:algo2}\>\>\`(\arabic{equation})
  \end{tabbing}

The available implementation \cite{CHC} provides both
$\Num{m}{n_1,\ldots,n_m}^{\prime}$ and
$\Num{m}{n_1,\ldots,n_m}^{\prime\prime}$, as well as direct
implementations of expansions of \eqref{eq:num}, like the ones presented in Eqs.~(\ref{eq:2pCorrelation})-(\ref{eq:4pCorrelation}), for all higher order correlators up to and including $m=8$. More
details about the implementation are available in
Appendix~\ref{s:Algorithm}.


\section{Monte Carlo studies}
\label{s:Monte Carlo studies}

In this section we illustrate with Monte Carlo studies how the generic framework outlined in previous sections can be used. Our exposition will branch into two main directions. Firstly, in a toy Monte Carlo study we illustrate how our framework can serve to correct for detector effects by working out two concrete examples which are regularly encountered as systematic biases in the anisotropic flow analyses. The first one is the systematic bias stemming from the non-uniform azimuthal detector acceptance. The second one is the systematic bias stemming from the non-uniform reconstruction efficiency as a function of transverse momentum. In order to correct for such effects, we will construct and use $\varphi$-weights and $p_{\rm T}$-weights, respectively. Secondly, in a realistic Monte Carlo study, we demonstrate how our framework can be used in the measurement of some new observables that we propose, and which were, with the techniques available so far experimentally inaccessible. We will conclude this section with estimates for these new observables in heavy-ion collisions at both  RHIC and LHC energies.  

We start by introducing the probability density function (p.d.f.), $f(\varphi)$, which will be used to sample the azimuthal angles of all particles. We consider $f(\varphi)$ to be a normalized Fourier-like p.d.f. parametrized with six harmonics $v_1, v_2, \ldots, v_6$, and the reaction plane $\Psi_{\rm RP}$. Written explicitly:
\begin{eqnarray}
f(\varphi) &=& \frac{1}{2\pi}\big[1+2v_1\cos(\varphi\!-\!\Psi_{\rm RP})+2v_2\cos(2(\varphi\!-\!\Psi_{\rm RP}))+2v_3\cos(3(\varphi\!-\!\Psi_{\rm RP}))\nonumber\\
&&{}+2v_4\cos(4(\varphi\!-\!\Psi_{\rm RP}))+2v_5\cos(5(\varphi\!-\!\Psi_{\rm RP}))+2v_6\cos(6(\varphi\!-\!\Psi_{\rm RP}))\big]\,.
\label{eq:pdf}
\end{eqnarray}
For each event we randomly determine the reaction plane $\Psi_{\rm RP}$ by uniformly sampling its value from an interval $\left[0,2\pi\right>$. Due to this randomization, which was directly motivated by random fluctuations in the direction of the impact parameter vector in real heavy-ion collisions, only the isotropic multi-particle correlators will have non-vanishing values once the data sample has been extended from a single event to multiple events~\cite{Bhalerao:2011yg}. In the above p.d.f. we assign to the flow harmonics the following input values:
\begin{eqnarray}
v_n = 0.04 +n\cdot 0.01,\qquad n = 1,2,\ldots, 6,
\label{eq:inputValues}
\end{eqnarray}
which are constant for all events. At first we set all six harmonics to be independent of transverse momentum and pseudorapidity, but we will relax this setting in the second part of this section when we allow the harmonic $v_2$ to have a non-trivial dependence on transverse momentum. Eq.~(\ref{eq:pdf}) then governs the distribution of the azimuthal angles of all particles, while the distribution of the other two kinematic variables, namely transverse momentum and pseudorapidity, are governed by the Boltzmann and uniform p.d.f.'s, respectively. For the Boltzmann p.d.f. we have used the following parametrization:
\begin{equation}
f(p_{\rm T}) = Mp_{\rm T}\exp\left(-\frac{\sqrt{m^2+p_{\rm T}^2}}{T}\right)\,,
\label{eq:Boltzmann}
\end{equation}
where $m$ is the mass of the particle, $T$ is the ``temperature'', and $M$ is the multiplicity of the event. We have set $m$ to be the mass of the charged pions, i.e. $m=0.13957$ GeV/$c^2$. By increasing the parameter $T$, one shifts the mean of the Boltzmann distribution towards higher $p_{\rm T}$ values, and  we have used $T=0.44$ GeV/$c$. In each event we have sampled precisely 500 particles, so as to avoid potential systematic biases due to trivial multiplicity fluctuations. Finally, we remark that in all separate toy MC studies we have set the random seed to be the same in order to isolate genuine systematic effects from trivial effects due to statistical fluctuations. 

We start with an example in which we illustrate how our formalism can be used to correct for systematic biases due to non-uniform acceptance in the azimuthal angles, after which we switch to an example that corrects for systematic biases due to non-uniform efficiency in particle reconstruction as a function of transverse momentum.


\subsection{$\varphi$-weights}
\label{ss:Phi-weights}

We select randomly one example for isotropic 2-, 3-, $\ldots$, and 8-p correlations, and, for simplicity, we use in this section a shorthand notation without subscripts for them. In particular, we have selected:
\begin{eqnarray}
\left<2\right> &\equiv& \left<2\right>_{-2,2} = v_2^2 = 3.6\times 10^{-3}\,,\nonumber\\
\left<3\right> &\equiv& \left<3\right>_{-5,-1,6} = v_1v_5v_6 = 4.5\times 10^{-4} \,,\nonumber\\
\left<4\right> &\equiv& \left<4\right>_{-3,-2,2,3} = v_2^2v_3^2 = 1.764\times 10^{-5} \,,\nonumber\\
\left<5\right> &\equiv& \left<5\right>_{-5,-4,3,3,3} = v_3^3v_4v_5 = 2.4696\times 10^{-6} \,,\nonumber\\
\left<6\right> &\equiv& \left<6\right>_{-2,-2,-1,-1,3,3} = v_1^2v_2^2v_3^2 = 4.41\times 10^{-8} \,,\nonumber\\
\left<7\right> &\equiv& \left<7\right>_{-6,-5,-1,1,2,3,6} = v_1^2v_2v_3v_5v_6^2 = 9.45\times 10^{-9} \,,\nonumber\\
\left<8\right> &\equiv& \left<8\right>_{-6,-6,-5,2,3,3,4,5} = v_2v_3^2v_4v_5^2v_6^2 = 1.90512 \times 10^{-9} \,.
\end{eqnarray}
Numerical values on the right-hand side in the above equations were obtained by calculating the theoretical values for each correlator from the Eq.~(\ref{eq:mixedHarmonicsExpVersion}), and inserting input values for flow harmonics from  (\ref{eq:inputValues}). We have rescaled observable $\left<k\right>$ by $10^{-k}$ in all figures, in order to plot all values on the same scale.  
%
\begin{figure}
\centering
\begin{minipage}{.5\textwidth}
  \centering
  \includegraphics[width=1.0\textwidth]{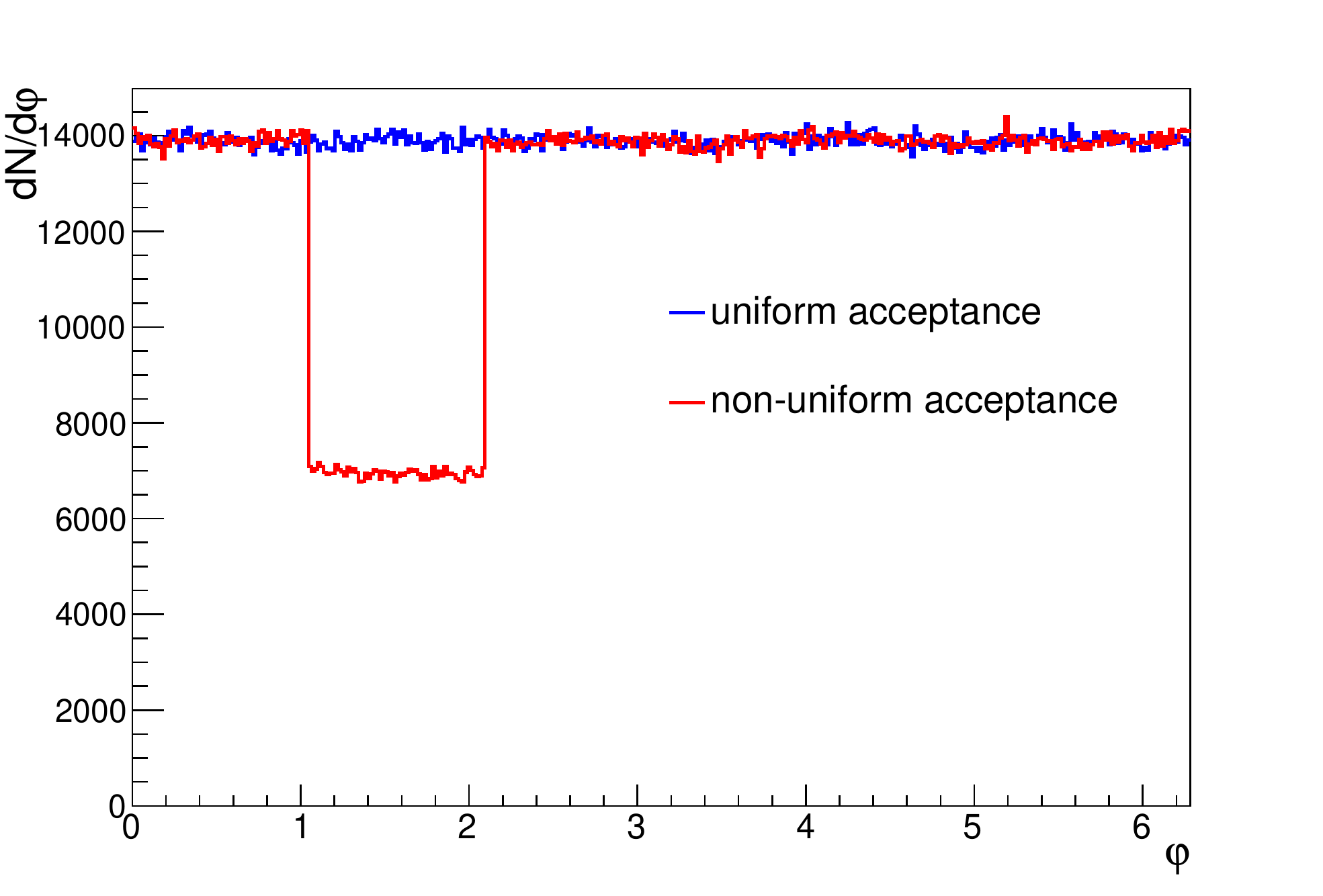}
  \caption{Azimuthal acceptance: uniform (blue)\\ and non-uniform (red), for a detector.}
  \label{fig:acceptance}
\end{minipage}%
\begin{minipage}{.5\textwidth}
  \centering
  \includegraphics[width=1.0\textwidth]{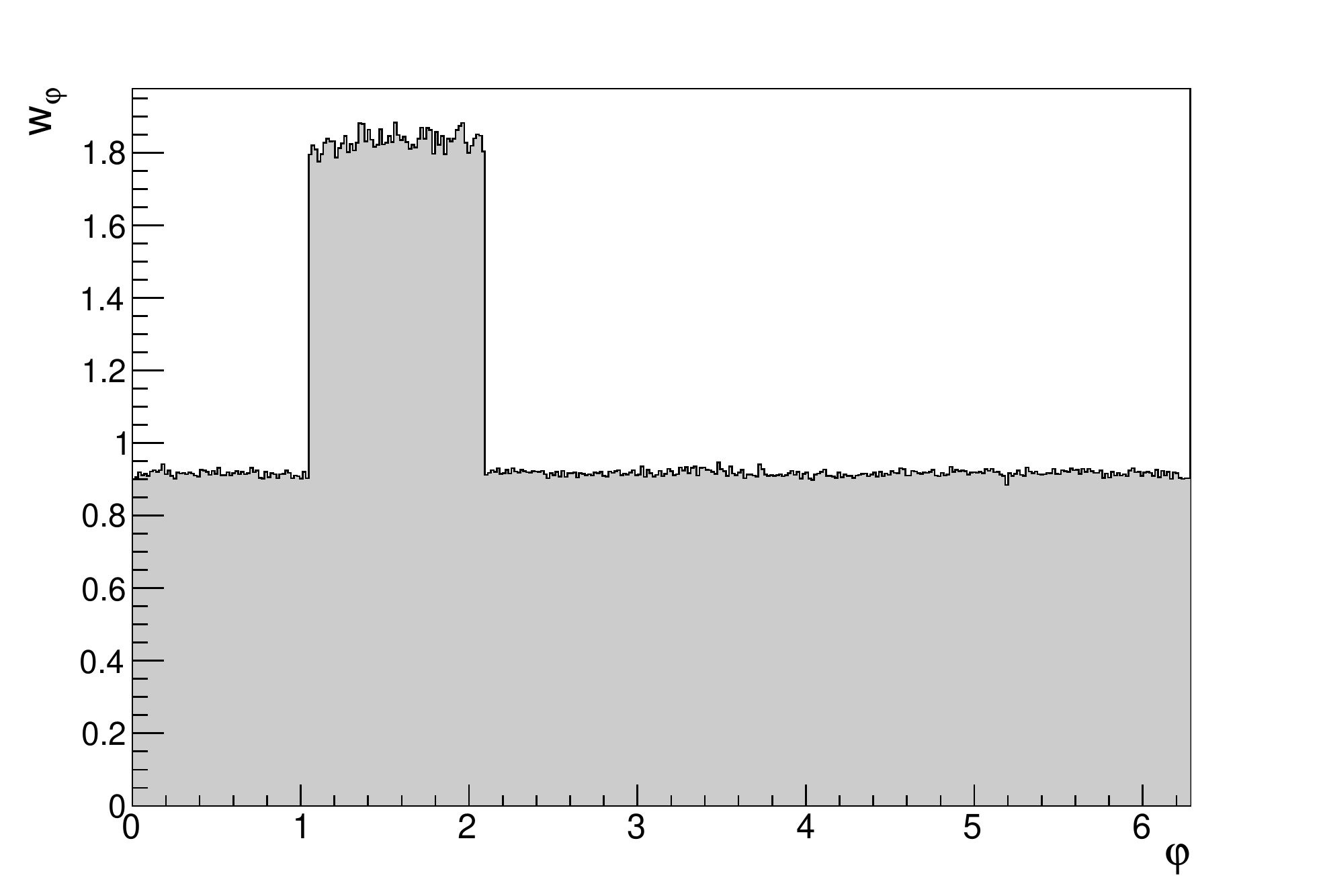}
  \caption{Resulting $\varphi$-weights for the case of non-uniform azimuthal acceptance shown in Fig.~\ref{fig:acceptance}.}
  \label{fig:phiWeights}
\end{minipage}
\end{figure}
\begin{figure}
\centering
\includegraphics[width=0.5\textwidth]{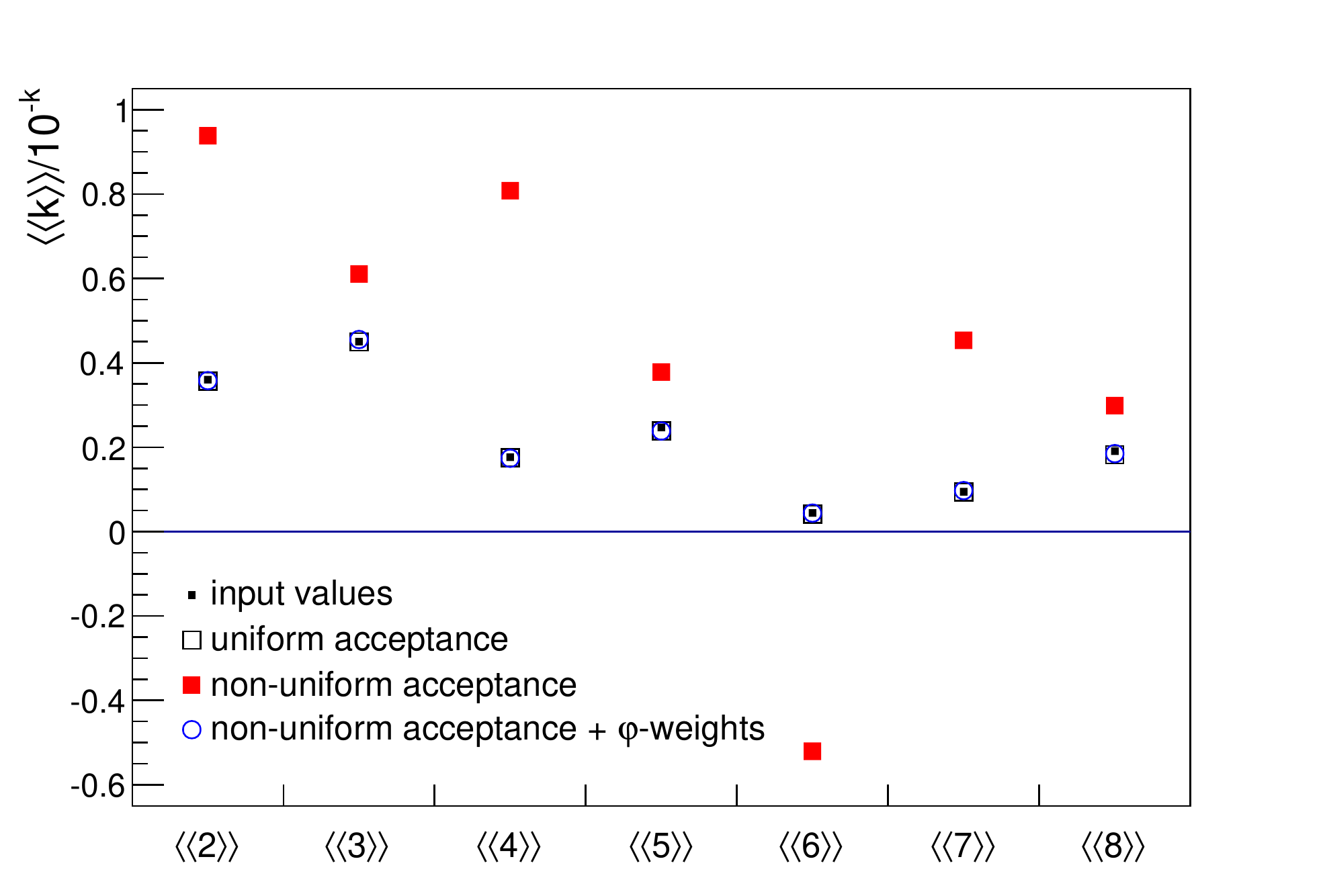}
\caption{Multi-particle observables corrected for non-uniform acceptance using  $\varphi$-weights compared to input values and values for uniform acceptance (see the text for the precise explanation of the ordinate.)}
\label{fig:phiResults}
\end{figure}
%

Our toy MC procedure consists of three separate runs. Firstly, we run our simulation    
for the case of uniform azimuthal acceptance, to demonstrate that the generic equations which we have derived reproduce correctly the input values for all multi-particle observables. This can be seen by comparing filled and open black markers in Fig.~\ref{fig:phiResults}. Secondly, we have rerun the simulation  using the same seed for random generation, but now have selected for analysis each particle with a probability which depends on its azimuthal angle. In particular, the particles which were sampled in the azimuthal range $60^{\rm o} \leq \varphi < 120^{\rm o}$ have been reduced by $50\%$ for this analysis. In this way we have simulated a non-uniform azimuthal detector acceptance (see Fig.~\ref{fig:acceptance}), and the corresponding non-negligible systematic bias in anisotropic flow analyses, which is depicted with red filled markers in Fig.~\ref{fig:phiResults}. In order to correct for this systematic bias, we have constructed $\varphi$-weights, $w_\varphi$, by inverting the histogram for non-uniform acceptance in Fig.~\ref{fig:acceptance}. The resulting $\varphi$-weights are shown in Fig.~\ref{fig:phiWeights}. We remark that in our framework the weights do not have to be normalized explicitly, because the analytic equations we provide for multi-particle correlations are normalized by definition (see Eq.~(\ref{eq:mpCorrelation})). Finally, we rerun the simulation for the third time with the same configuration as in the second run, now utilizing the constructed $\varphi$-weights from Fig.~\ref{fig:phiWeights} when we are filling $Q$-vectors~(\ref{eq:Qvector}) in each event. As can be seen from the blue open circles in Fig.~\ref{fig:phiResults}, $\varphi$-weights completely suppress the systematic bias from non-uniform acceptance for all multi-particle observables we have selected in this example. 
 
Based on the previous example, we conclude that as far as $\varphi$-weights can be constructed for the measured azimuthal distribution, our generic framework can be used to correct for the systematic bias for the cases when that distribution is non-uniform, and it is applicable for any multi-particle observable even when multiple harmonics are present in the system. These two points improve and generalize the prescription outlined in Appendix B of~\cite{Bilandzic:2010jr}. In the next example, we will demonstrate the usage of $p_{\rm T}$-weights.

\subsection{$p_{\rm T}$-weights}
\label{ss:Pt-weights}

In this part of the study we use the same MC setup established in the previous example for the $\varphi$-weights with one exception. In this example we introduce the following $p_{\rm T}$ dependence of $v_2$:
\begin{equation}
v_2(p_{\rm T}) = \left\{
\begin{array}{ll}
v_{2,\rm max}(p_{\rm T}/p_{\rm cutoff}) & p_{\rm T} < p_{\rm cutoff} \\
v_{2,\rm max} & p_{\rm T} \geq p_{\rm cutoff}\,, \\
\end{array}
\right.
\label{eq:v(pt)parametrization}
\end{equation}
and we have set the above parameters to $p_{\rm cutoff} = 2.0$ GeV/$c$ and $v_{2,\rm max} = 0.3$. 

Again, we have randomly selected one example for isotropic 2-, 3-, $\ldots$, and 8-p correlations (suppressing their subscripts for simplicity in the rest of this section):
\begin{eqnarray}
\left<2\right> &\equiv& \left<2\right>_{-2,2} = v_2^2\,,\nonumber\\
\left<3\right> &\equiv& \left<3\right>_{-5,-1,6} = v_1v_5v_6\,,\nonumber\\
\left<4\right> &\equiv& \left<4\right>_{-5,-2,2,5} = v_2^2v_5^2\,,\nonumber\\
\left<5\right> &\equiv& \left<5\right>_{-5,-4,-1,4,6} = v_1v_4^2v_5v_6\,,\nonumber\\
\left<6\right> &\equiv& \left<6\right>_{-2,-2,-2,-2,3,5} = v_2^4v_3v_5\,,\nonumber\\
\left<7\right> &\equiv& \left<7\right>_{-2,-2,-2,-1,2,2,3} = v_1v_2^5v_3\,,\nonumber\\
\left<8\right> &\equiv& \left<8\right>_{-5,-4,-2,-2,2,2,4,5} = v_2^4v_4^2v_5^2\,.
\label{eq:choicePt}
\end{eqnarray}
Some of the selected observables ($\left<\left<3\right>\right>$ and $\left<\left<5\right>\right>$) do not have an explicit dependence on $v_2$, so we do not expect them to exhibit any systematic bias in this example. 

\begin{figure}
\centering
\begin{minipage}{.5\textwidth}
  \centering
  \includegraphics[width=1.0\textwidth]{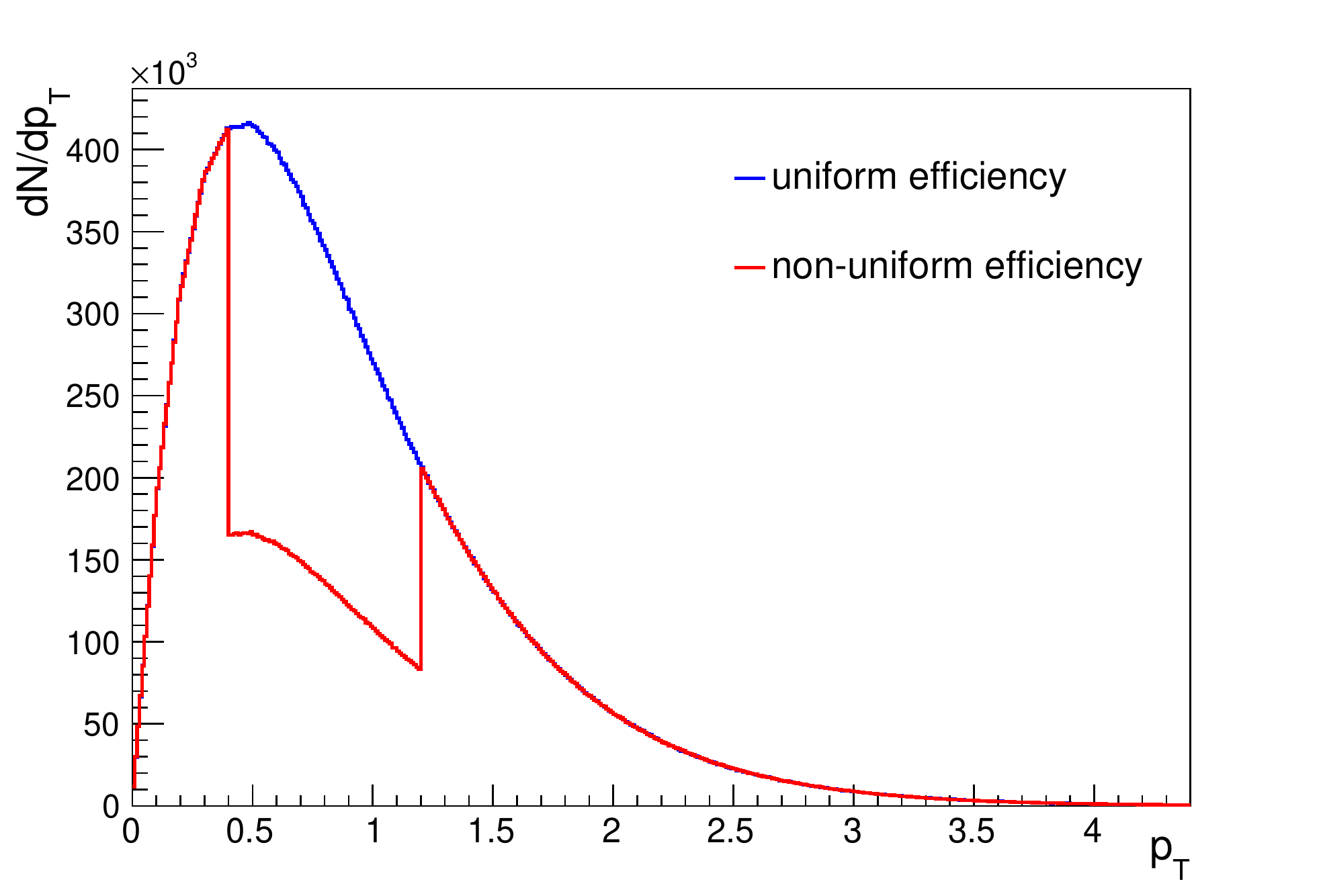}
  \caption{Transverse momentum yield, with uniform (blue)\\ and non-uniform (red) reconstruction efficiency.}
  \label{fig:efficiency}
\end{minipage}%
\begin{minipage}{.5\textwidth}
  \centering
  \includegraphics[width=1.0\textwidth]{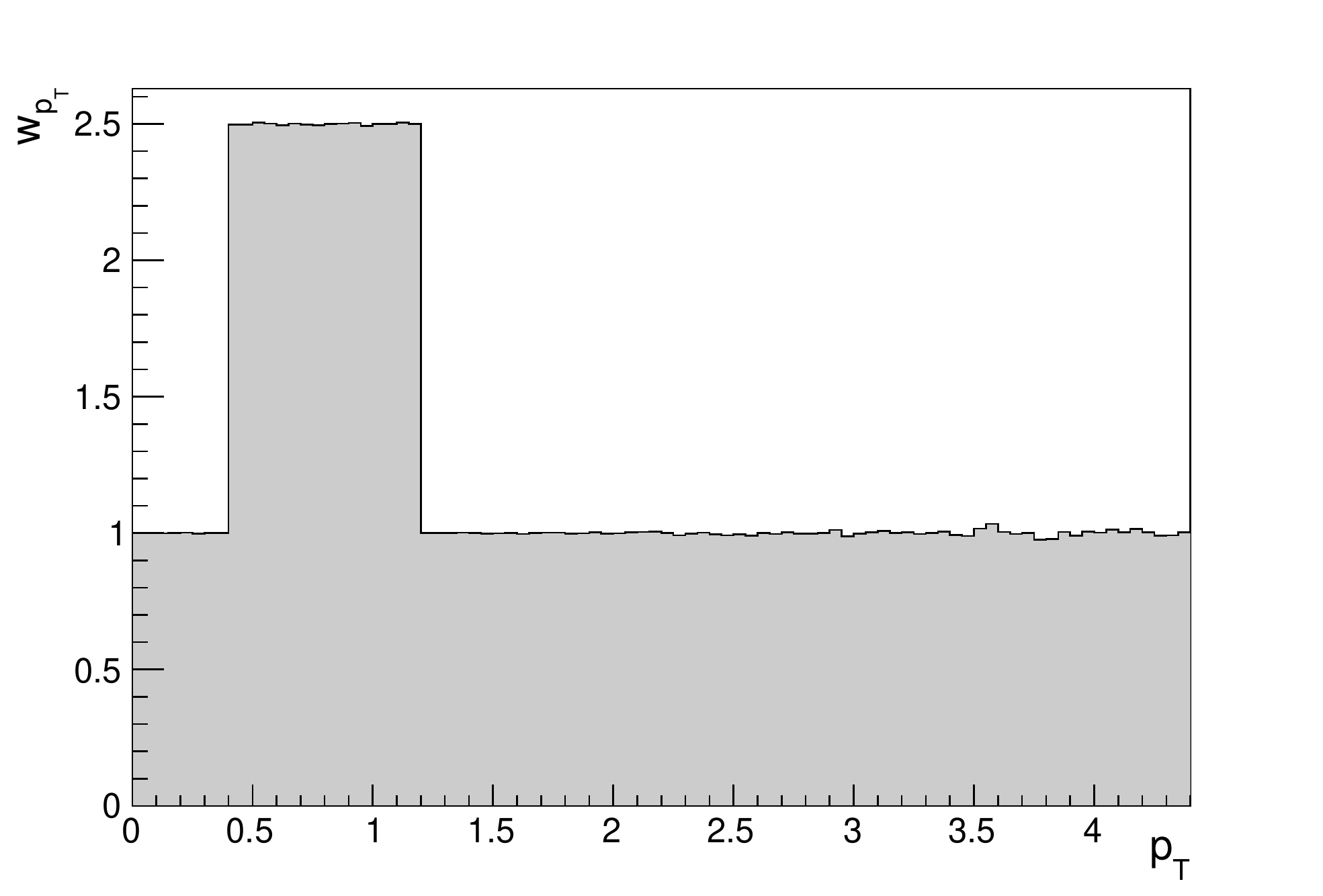}
  \caption{Resulting $p_{\rm T}$-weights corresponding to the non-uniform efficiency shown in Fig.~\ref{fig:efficiency}.}
  \label{fig:ptWeights}
\end{minipage}
\end{figure}

\begin{figure}
\centering
\includegraphics[width=0.5\textwidth]{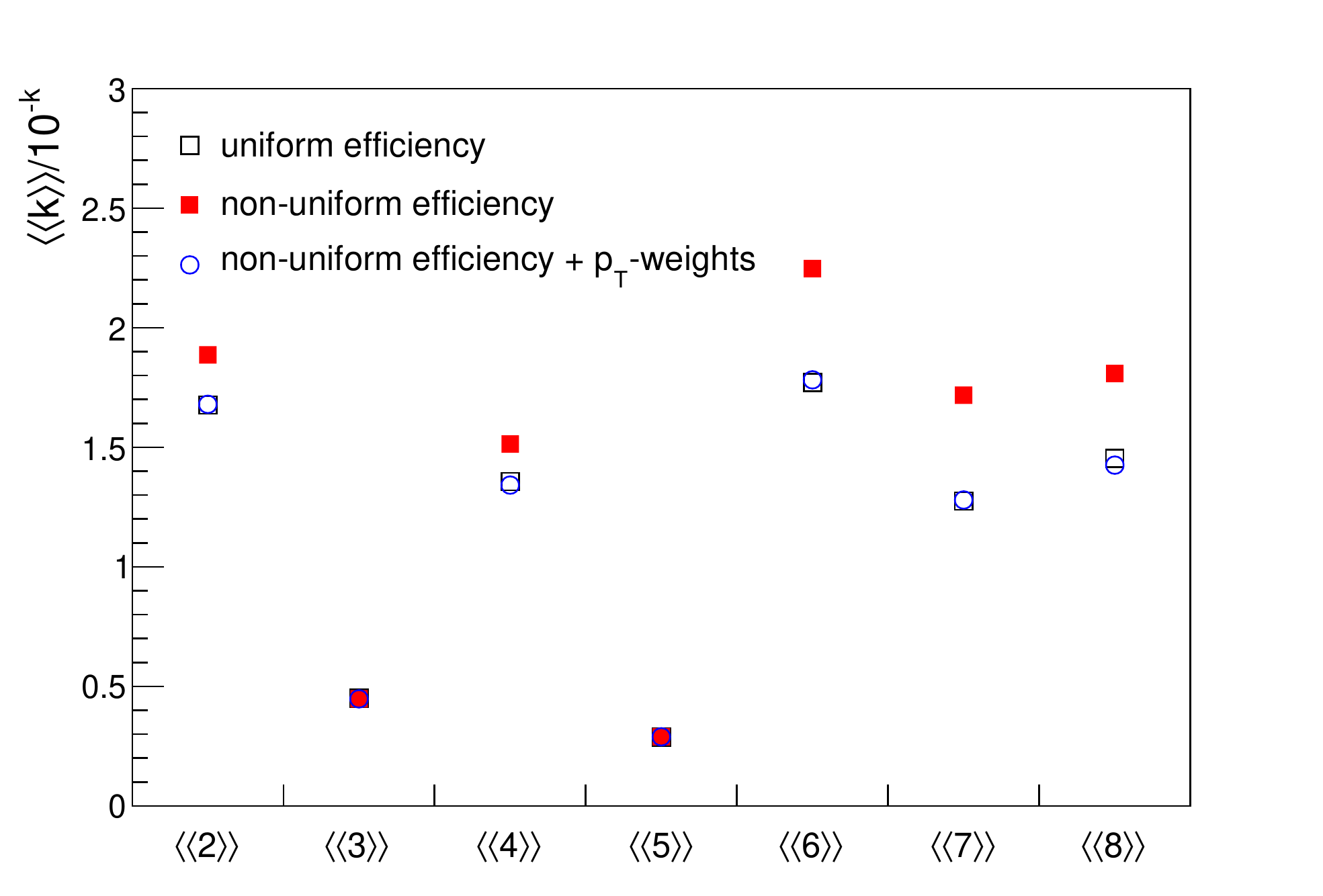}
\caption{Results from a toy MC simulation for multi-particle observables corrected for non-uniform efficiency using $p_{\rm T}$-weights.}
\label{fig:ptResults}
\end{figure}

Analogously as in the previous example, our toy MC procedure consists of three separate runs. Firstly, we run our simulation for the case of uniform reconstruction efficiency, in order to obtain the true $p_{\rm T}$ yield; this result is illustrated with the blue line in Fig.~\ref{fig:efficiency}. Secondly, we have rerun the same simulation, but now have selected for the analysis each particle with a probability which depends on its transverse momentum. The particles in the  
transverse momentum interval $0.4 \leq p_{\rm T} < 1.2$ have been reduced by $60\%$. The resulting $p_{\rm T}$ yield is depicted by the red line in Fig.~\ref{fig:efficiency}. The resulting systematic bias on the selected multi-particle observables~(\ref{eq:choicePt}) can be seen by inspecting the red filled markers in Fig.~\ref{fig:ptResults}. As already remarked, such a bias is absent in observables $\left<\left<3\right>\right>$ and $\left<\left<5\right>\right>$, because they do not have the explicit dependence on the harmonic $v_2$ (see (\ref{eq:choicePt})), which is the only harmonic in this study which has a non-trivial $p_{\rm T}$ dependence. To correct for reconstruction efficiency,  we have constructed $p_{\rm T}$-weights, $w_{p_{\rm T}}$, by taking the ratio of the two histograms in Fig.~\ref{fig:efficiency}. The result is shown in Fig.~\ref{fig:ptWeights}. Finally, in the third run we use the same MC setup as in the second run, only now we make use of the constructed $p_{\rm T}$-weights from Fig.~\ref{fig:ptWeights} when filling the $Q$-vectors~(\ref{eq:Qvector}). The agreement between the results shown with black open squares (uniform efficiency) and the ones shown with blue open circles (non-uniform efficiency using the $p_{\rm T}$-weights) in Fig.~\ref{fig:ptResults}, demonstrates clearly that the generic framework is capable of suppressing the systematic bias from non-uniform efficiency for all of the multi-particle observables in question. 
%
%

With the previous two examples we have demonstrated that, in a simple toy MC study, our generic framework can be utilized to correct for various detector inefficiencies. Next, we will illustrate, in a study based on a realistic MC model, its use in the measurement of some new physical observables which we now propose.  

\subsection{Example new observables: ``Standard candles"}
\label{ss:Example new observables: ``Standard candles"}

We now introduce a new type of observable for anisotropic flow analyses, the so-called \textit{standard candles (SC)}, which can be measured with the generic framework we have presented in the previous sections. This observable is particularly useful for systems in which flow harmonics  fluctuate in magnitude event-by-event (the case we have in reality). We start with the following generic four-particle correlation:
\begin{equation}
\left<\left<\cos(m\varphi_1\!+\!n\varphi_2\!-\!m\varphi_3-\!n\varphi_4)\right>\right>\,,
\label{eq:4p_sc_correlation}
\end{equation}
and we impose the constraint $m\neq n$. The isotropic part of corresponding four-particle cumulant is given by:
\begin{eqnarray}
\left<\left<\cos(m\varphi_1\!+\!n\varphi_2\!-\!m\varphi_3-\!n\varphi_4)\right>\right>_c &=& \left<\left<\cos(m\varphi_1\!+\!n\varphi_2\!-\!m\varphi_3-\!n\varphi_4)\right>\right>\nonumber\\
&&{}-\left<\left<\cos[m(\varphi_1\!-\!\varphi_2)]\right>\right>\left<\left<\cos[n(\varphi_1\!-\!\varphi_2)]\right>\right>\nonumber\\
&=&\left<v_{m}^2v_{n}^2\right>-\left<v_{m}^2\right>\left<v_{n}^2\right>\nonumber\\
&=&0\,,
\label{eq:4p_sc_cumulant}
\end{eqnarray}
where double angular brackets indicate that the averaging from definition~(\ref{eq:mpCorrelation}) has been extended to all events. Due to the condition that $m\neq n$, a lot of terms which appear in the general cumulant expansion, for instance $\left<\left<\cos(m\varphi_1\!-\!n\varphi_2)\right>\right>$, are non-isotropic and, therefore, average to zero for a detector with uniform acceptance when the averaging is extended to all events. For fixed values of $v_m$ and $v_n$ over all events, the four-particle cumulant as defined in Eq.~(\ref{eq:4p_sc_cumulant}), is zero by definition. Any dependence on the symmetry planes $\Psi_m$ and $\Psi_n$ is also canceled by definition. We can get the result in the last line of Eq.~(\ref{eq:4p_sc_cumulant}) not only when $v_m$ and $v_n$ are fixed for all events, but also when event-by-event fluctuations of $v_m$ and $v_n$ are uncorrelated, since the expression $\left<v_{m}^2v_{n}^2\right>$ can then be factorized.
Taking all these statements into account, the four-particle cumulant~(\ref{eq:4p_sc_cumulant}) is non-zero only if the event-by-event fluctuations of $v_m$ and $v_n$ are correlated. Therefore, by measuring the observable~(\ref{eq:4p_sc_cumulant}) we can conclude whether finding $v_m$ larger than $\left<v_m\right>$ in an event will enhance or reduce the probability of finding $v_n$ larger than $\left<v_n\right>$ in that event, which is not constrained by any measurement performed yet. Since by definition everything cancels out from the observable~(\ref{eq:4p_sc_cumulant}) except the last contribution, namely the correlation of event-by-event fluctuations of $v_m$ and $v_n$, we name it a ``standard candle". Recently, by using different observables and methodology, these correlations between fluctuations of various harmonics have been studied in~\cite{Niemi:2012aj,Huo:2013qma}.

In this study, the Monte Carlo event generator, A MultiPhase Transport (AMPT) model, has been used. AMPT is a hybrid model consisting of four main parts: the initial conditions, partonic interactions, hadronization, and hadronic rescatterings. The initial conditions, which include the spatial and momentum distributions of minijet partons and soft string excitations, are obtained from the Heavy Ion Jet Interaction Generator (HIJING)~\cite{Wang:2000bf}. 
The following stage which describes the interactions between partons is modeled by Zhang's Parton Cascade (ZPC)~\cite{Zhang:1997ej}, which presently includes only two--body scatterings with cross sections obtained from pQCD with screening masses. In AMPT with string melting~\cite{Lin:2003iq}, the transition from partonic to hadronic matter is done through a simple coalescence model, which combines two quarks into mesons and three quarks into baryons~\cite{Chen:2005mr}. To describe the dynamics of the subsequent hadronic stage, a hadronic cascade, which is based on A Relativistic Transport (ART) model~\cite{Li:1995pra}, is used.

Several configurations of the AMPT model have been investigated to better understand the results based on AMPT simulations~\cite{Zhou:2010us}. 
The partonic interactions can be tweaked by changing the partonic cross section: for RHIC the default value is 10 mb, while using 3 mb generates weaker partonic interactions in ZPC. We can also change the hadronic interactions by controlling the termination time in ART.
Setting NTMAX = 3 will turn off the hadronic interactions effectively~\cite{Zhou:2010us}. Good agreement has been observed recently between anisotropic flow measurements and the AMPT~\cite{Xu:2011fi}. Therefore, we calculate multi-particle azimuthal correlations using AMPT simulations with the input parameters suggested
in~\cite{Xu:2011fi} at the LHC energy. For RHIC energies we followed the parameters in~\cite{Chen:2005mr} while different configurations have also been used in this study.

\begin{figure}
\centering
\includegraphics[width=0.5\textwidth]{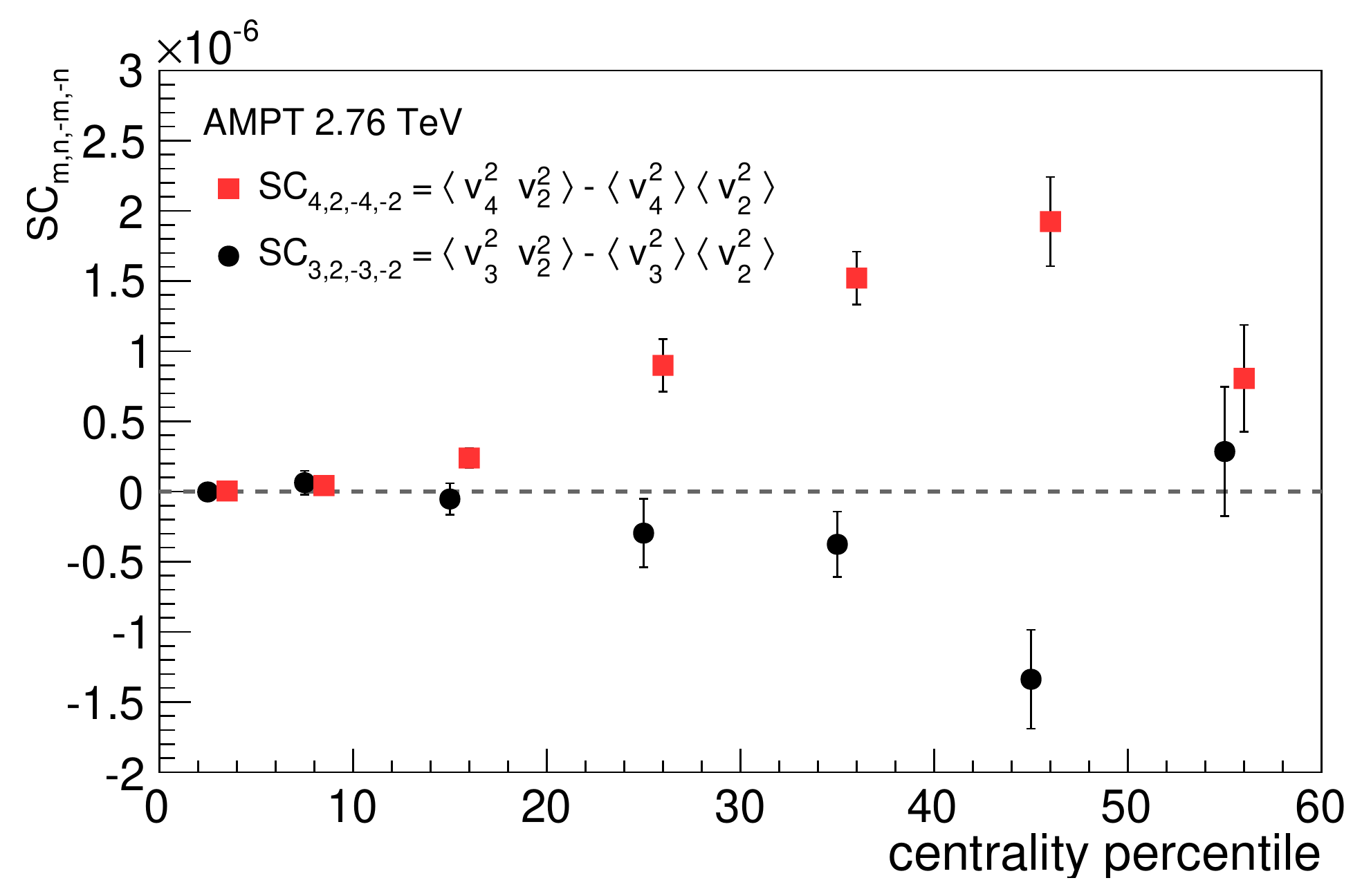}
\caption{The centrality dependence of standard candles $SC_{4,2,-4,-2}$ (red markers) and $SC_{3,2,-3,-2}$ (black markers) at $\sqrt{s_{\rm NN}}$ = 2.76 TeV Pb--Pb collisions with AMPT-StringMelting.}
\label{fig:sc}
\end{figure}

In Fig.~\ref{fig:sc} we see a clear non-zero value for both $SC_{4,2,-4,-2}$ (red markers) and $SC_{3,2,-3,-2}$ (black markers) at the LHC energy. The positive results of $SC_{4,2,-4,-2}$ suggest a positive correlation between the event-by-event fluctuations of $v_{2}$ and $v_{4}$, which indicates that finding $v_{2}$ larger than $\langle v_{2} \rangle$ in an event enhances the probability of finding $v_{4}$ larger than $\langle v_{4} \rangle$ in that event. On the other hand, the negative results of $SC_{3,2,-3,-2}$ predict that finding $v_{2}$ larger than $\langle v_{2} \rangle$ enhances the probability of finding $v_{3}$ smaller than $\langle v_{3} \rangle$. 

A similar centrality dependence of $SC_{4,2,-4,-2}$ and $SC_{3,2,-3,-2}$ is also  found at the RHIC energy, see Fig.~8. In addition, we compare the $SC_{4,2,-4,-2}$ and $SC_{3,2,-3,-2}$ calculations for three different scenarios: (a) 3 mb; (b) 10 mb; (c) 10 mb, no rescattering.  
It was shown~\cite{Zhou:2013} that the relative flow fluctuations of $v_2$ do not depend on the partonic interactions and only relate to the initial eccentricity fluctuations. Therefore, the expectation is that $SC_{4,2,-4,-2}$ and $SC_{3,2,-3,-2}$ do not depend on the magnitudes of $v_{2}$ or $v_{4}$ (which depend on both partonic interactions and hadronic interactions), but depend only on the initial correlations of event-by-event fluctuations of $\varepsilon_{2}$ and $\varepsilon_{4}$. 
Thus, both $SC_{4,2,-4,-2}$ and $SC_{3,2,-3,-2}$ remain the same for different configurations, since the initial state was kept the same each time.
However, we find that when  the partonic cross section is decreasing from 10 mb (lower shear viscosity, see~\cite{Xu:2011fi}) to 3 mb (higher shear viscosity), the strength of $SC_{4,2,-4,-2}$ decreases. Additionally, the `10mb, no rescattering' setup seems to give slightly smaller magnitudes of $SC_{4,2,-4,-2}$ and $SC_{3,2,-3,-2}$. 

Considering the AMPT model can quantitatively describe flow measurements at the LHC~\cite{Xu:2011fi}, our AMPT calculations for these new observables provide predictions for the correlations of event-by-event fluctuations of $v_{2}$ and $v_{4}$, and of $v_{2}$ and $v_{3}$ for the measurements at the LHC. Such measurements have the potential to shed new light on the underlying physical mechanisms behind flow fluctuations.

\begin{figure}
\centering
\includegraphics[width=0.5\textwidth]{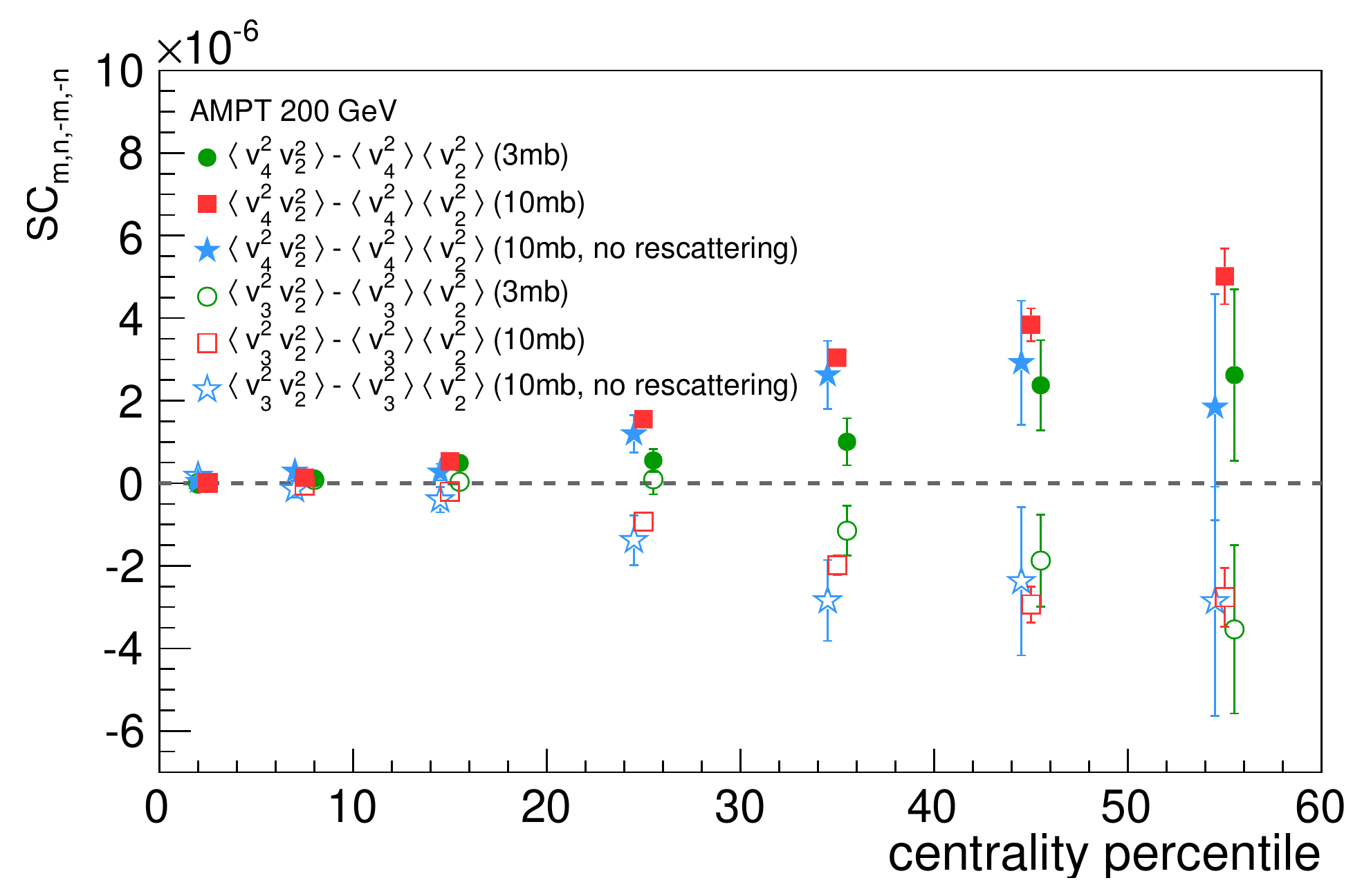}
\caption{ The centrality dependence of standard candles $SC_{4,2,-4,-2}$ (solid markers) and $SC_{3,2,-3,-2}$ (open markers) at $\sqrt{s_{\rm NN}} = $ 200 GeV 
Au--Au collisions with AMPT-StringMelting. Different scenarios: (a) 3 mb (green circle); (b) 10 mb (red square) and (c) 10 mb, no rescattering (azure star) are 
presented. }
\label{fig:sc_rhic}
\end{figure}
%


\section{Detectors with finite granularity}
\label{s:Detectors with finite granularity}

The previous results and examples are applicable only directly to detectors that have infinite resolution. Finite resolution
will both bias measurements and cause interference between harmonics. To study this, we
define a detector with $N$ equal size adjacent azimuthal sectors where the edge of the first
sector is shifted from 0 by $\varphi_\Delta$. Then the low and high edges of the
$i^{\mathrm{th}}$ sector are defined as follows:
\begin{equation}
\varphi_{L_{i}} = i\frac{2\pi}{N}+\varphi_\Delta,\qquad \varphi_{H_{i}} = (i+1)\frac{2\pi}{N}+\varphi_\Delta,\qquad 0 \leq i \leq N-1\,.
\end{equation}
By integrating Eq.~(\ref{eq:Fourier}) between $\varphi_{L_{i}}$ and $\varphi_{H_{i}}$ (derivation is shown in Appendix~\ref{s:Appendix_to_finite_granularity}), the probability, $P_i$, 
for a particle to be detected in the $i^{\mathrm{th}}$ sector is found to be:
\begin{equation}
P_i = \frac{1}{N}\left[1+\sum_{n=1}^{\infty}2v_n\frac{\sin\frac{n\pi}{N}}
{\frac{n\pi}{N}}\cos\left(n\bigg((i+\frac{1}{2})\frac{2\pi}{N}+\varphi_\Delta-\Psi_n\bigg)\right)\right]\,.
\label{eq:Pi}
\end{equation}
The expectation value for an observable for a single particle, $\theta$, can then be evaluated from $P_i$ using
the following formula:
\begin{equation}
E\left[\theta\right] = \sum_{i=0}^{N-1}\theta_i P_i\,,
\end{equation}
where $\theta_i$ is the value of observable evaluated at the center of the sector. It follows that the expectation value of $e^{in\varphi}$ (see derivation in Appendix~\ref{s:Appendix_to_finite_granularity}) is given by:
\begin{equation}
E\left[e^{in\varphi}\right] = \left\{
\begin{tabular}{ll}
$(-1)^{\frac{n}{N}} e^{in\varphi_\Delta}$ & for $\frac{n}{N} \in \mathbb{Z}$\\
$\sum\limits_{j=-\infty}^{\infty} v_{\left|jN-n\right|}\frac{\sin(j-\frac{n}{N})\pi}{(j-\frac{n}{N})\pi}(-1)^j e^{-i\left\{(jN-n)\Psi_{\left|jN-n\right|}-jN\varphi_{\Delta}\right\}}$ & for $\frac{n}{N} \notin \mathbb{Z}$
\end{tabular}
\right.
\label{eq:harmonic_expectation}
\end{equation}
where $\mathbb{Z}$ is the set of all integers. It is evident from this formula that it is not possible to measure any
harmonic which is a multiple of the number of sectors. If $0 < n < N/2$ and the harmonics above $N/2$ can be neglected, Eq.~(\ref{eq:harmonic_expectation}) becomes:
\begin{equation}
E\left[e^{in\varphi}\right] \approx
v_{n} e^{in\Psi_n} \frac{\sin\frac{n}{N}\pi}{\frac{n}{N}\pi}\,.
\end{equation}
In this case, the multi-particle azimuthal correlations defined in Eq.~(\ref{eq:mpCorrelation}) become (under the assumption (\ref{eq:factorization}) of a factorizable p.d.f.):
\begin{equation}
E\left[\left<m\right>_{n_1,\ldots,n_m}\right] \approx \prod_{k=1}^m
v_{n_k} e^{in_k\Psi_{n_k}} \frac{\sin\frac{n_k}{N}\pi}{\frac{n_k}{N}\pi}\,.
\label{eq:product}
\end{equation}
In this way, the term $\left(\frac{n}{N}\pi\right)/\sin\left(\frac{n}{N}\pi\right)$ is a correction
factor for a bias from finite granularity that must be applied for each harmonic that the multi-particle correlator
is composed of due to an overall reduction in the measured value.

\begin{figure}
\centering
\begin{minipage}{.5\textwidth}
  \centering
  \includegraphics[width=1.0\textwidth]{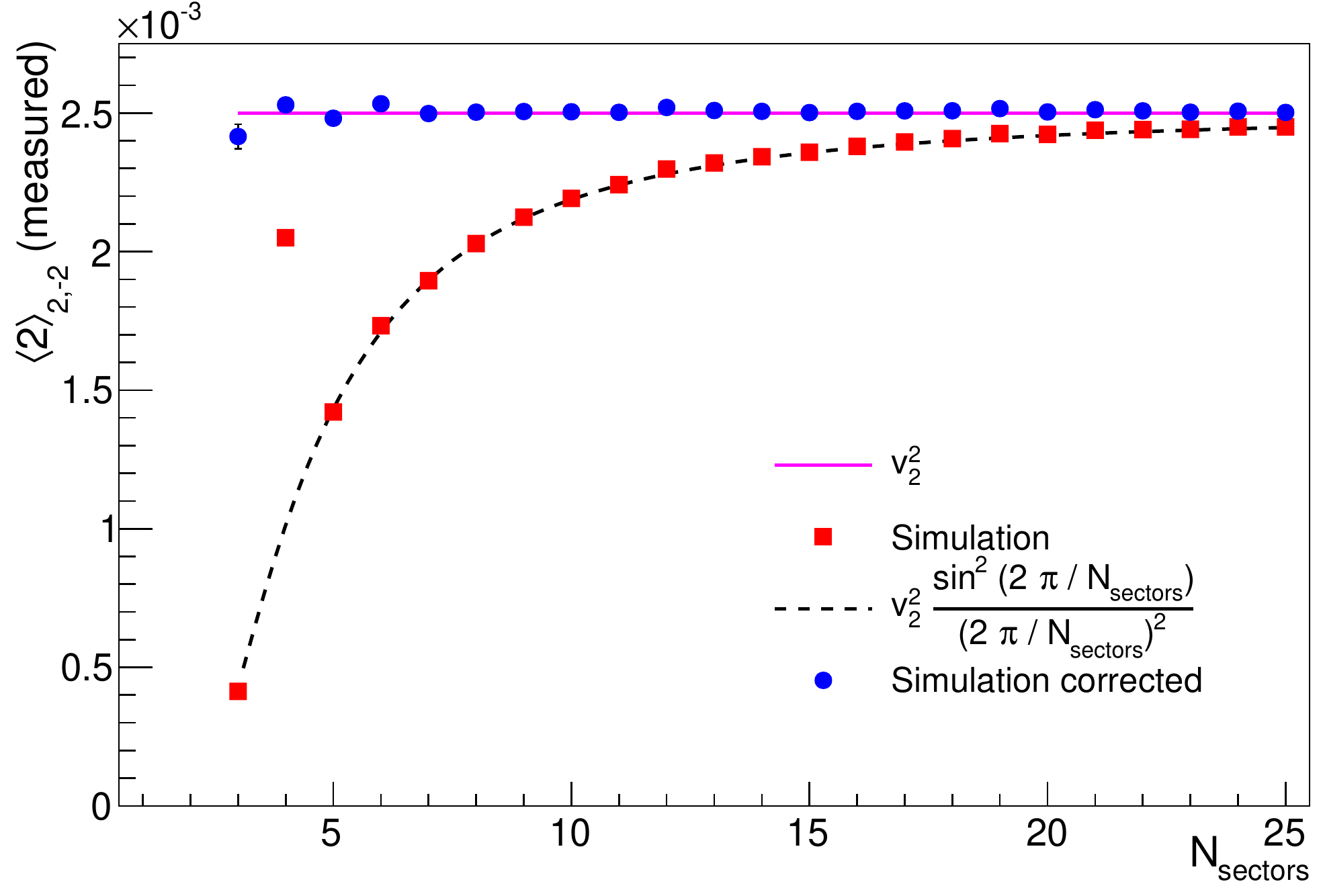}
\end{minipage}%
\begin{minipage}{.5\textwidth}
  \centering
  \includegraphics[width=1.0\textwidth]{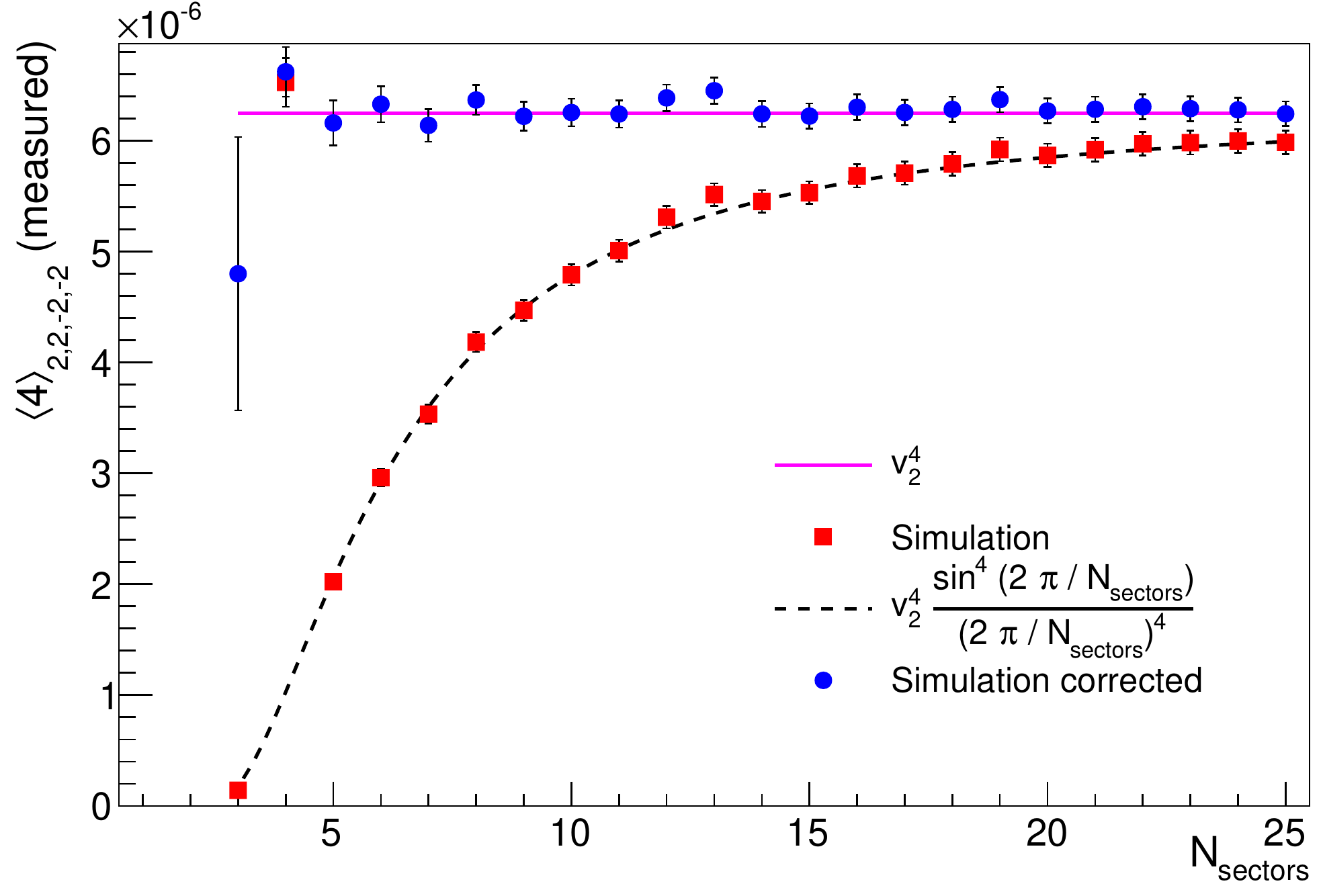}
\end{minipage}
\caption{$\langle 2 \rangle_{2,-2}$ and $\langle 4 \rangle_{2,2,-2,-2}$ (left and right, respectively) evaluated
for a range of sectors when only $v_{2}$ exists (red squares). The magenta line shows the input value of $v_{2}^{2}$ or $v_{2}^{4}$.
The black dashed line shows the expected measured value of the correlator. The blue circles are the simulated values
corrected for the reduction to the measured value due to finite granularity.}
\label{fig:v2vsnsectors}
\end{figure}

Figure~\ref{fig:v2vsnsectors} shows the result obtained by calculating
$\langle 2 \rangle_{2,-2}$ and $\langle 4 \rangle_{2,2,-2,-2}$ for detectors with various segmentations, and when the toy Fourier-like p.d.f. was parametrized only with the single harmonic $v_2$. The
simulated values lie on the dashed line suppressed by 2 or 4 factors of $\sin\left(\frac{2}{N}\pi\right)/\left(\frac{2}{N}\pi\right)$ (see Eq.~(\ref{eq:product})).
In this case (if $N > 2$), the values can be corrected to reproduce the input values of $v_{2}^{2}$ or $v_{2}^{4}$. The `blip'
at $N = 4$ is a special case where multiple factors proportional to $v_{2}$ in Eq.~(\ref{eq:harmonic_expectation})
contribute making the measured value 2 times bigger for $\langle 2 \rangle_{2,-2}$ and 6 times bigger for $\langle 4 \rangle_{2,2,-2,-2}$
than the expected suppressed value (the black dashed line) when averaging over all events.

\begin{figure}
\centering
\begin{minipage}{.5\textwidth}
  \centering
  \includegraphics[width=1.0\textwidth]{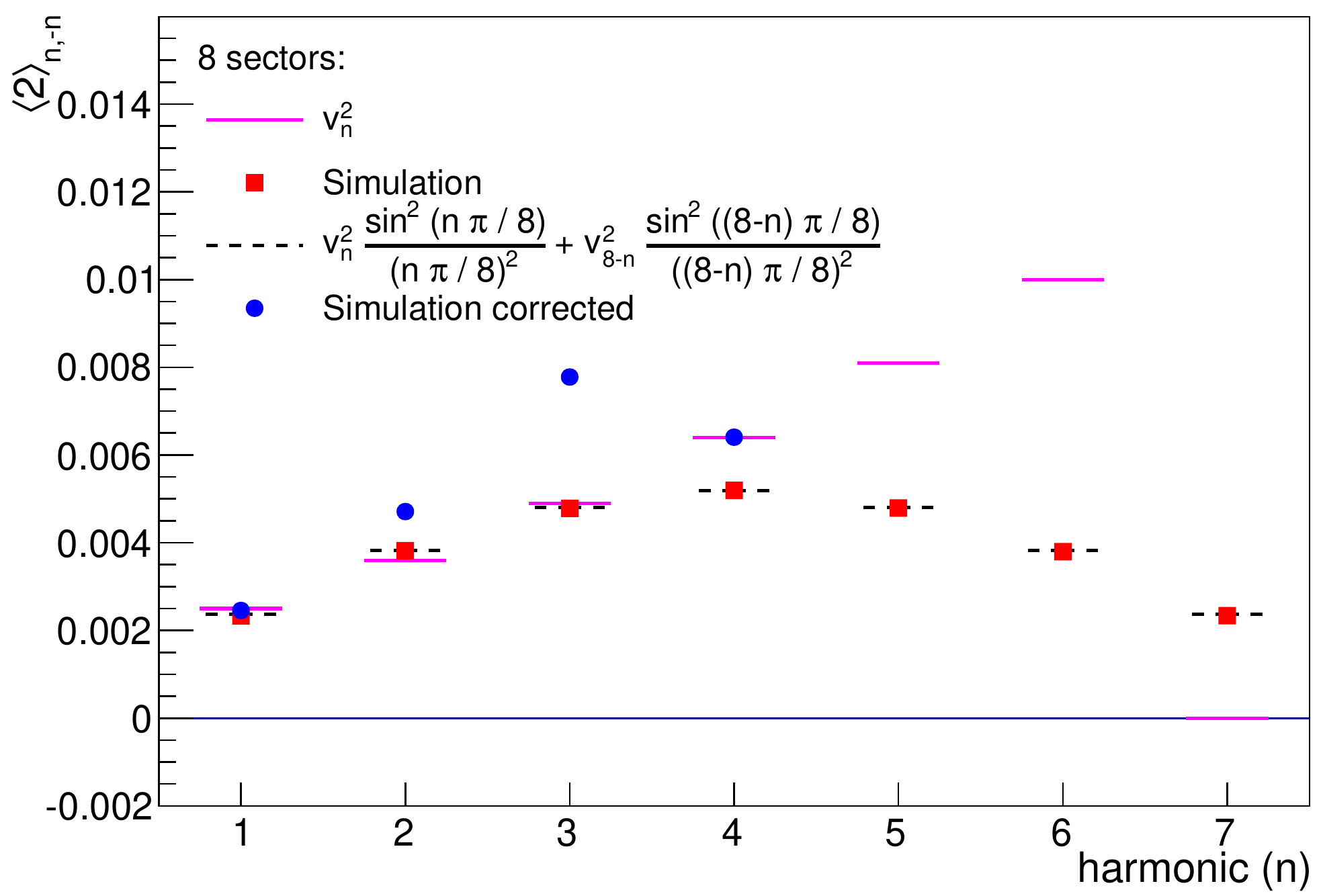}
\end{minipage}%
\begin{minipage}{.5\textwidth}
  \centering
  \includegraphics[width=1.0\textwidth]{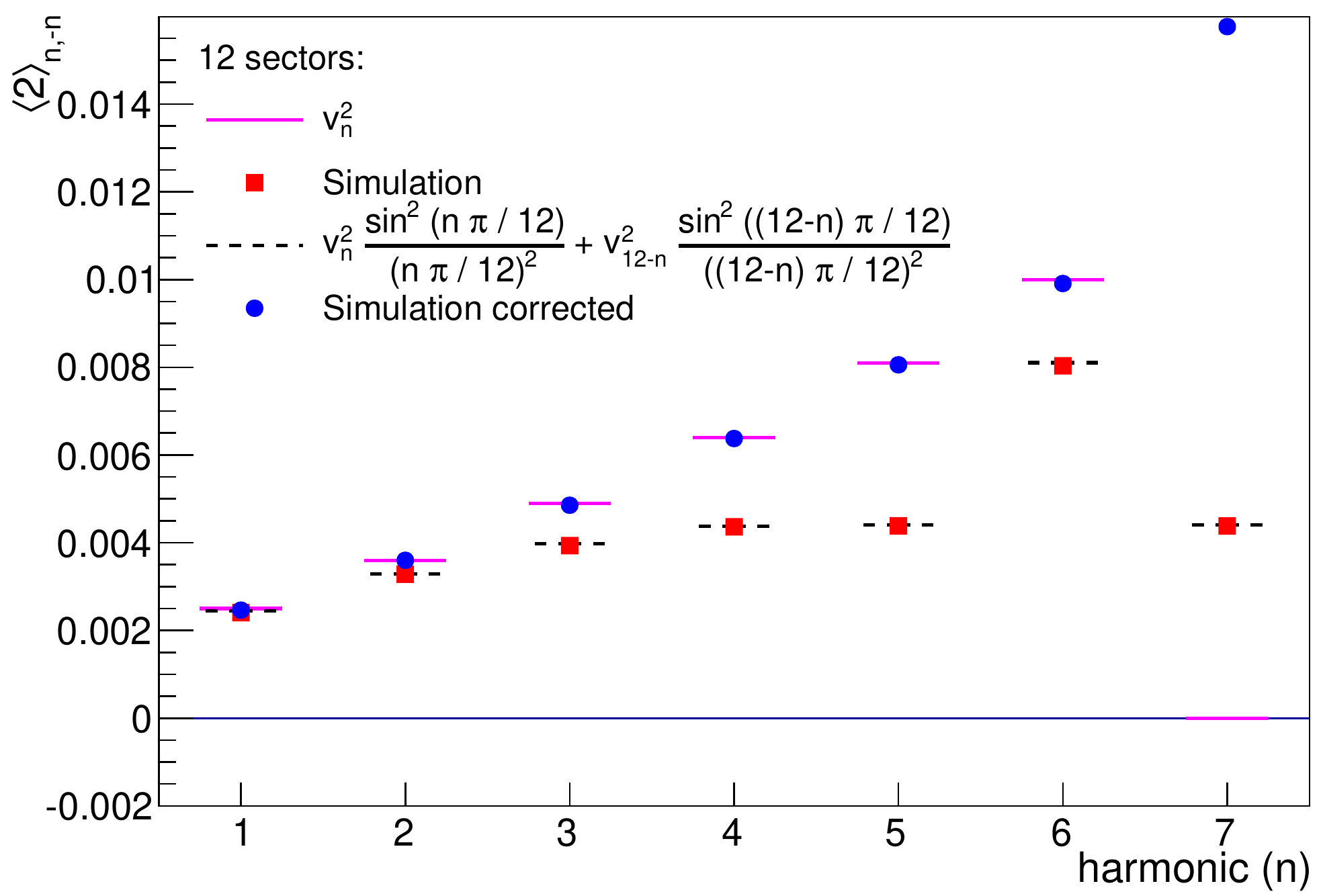}
\end{minipage}
\begin{minipage}{\textwidth}
  \centering
  \includegraphics[width=0.5\textwidth]{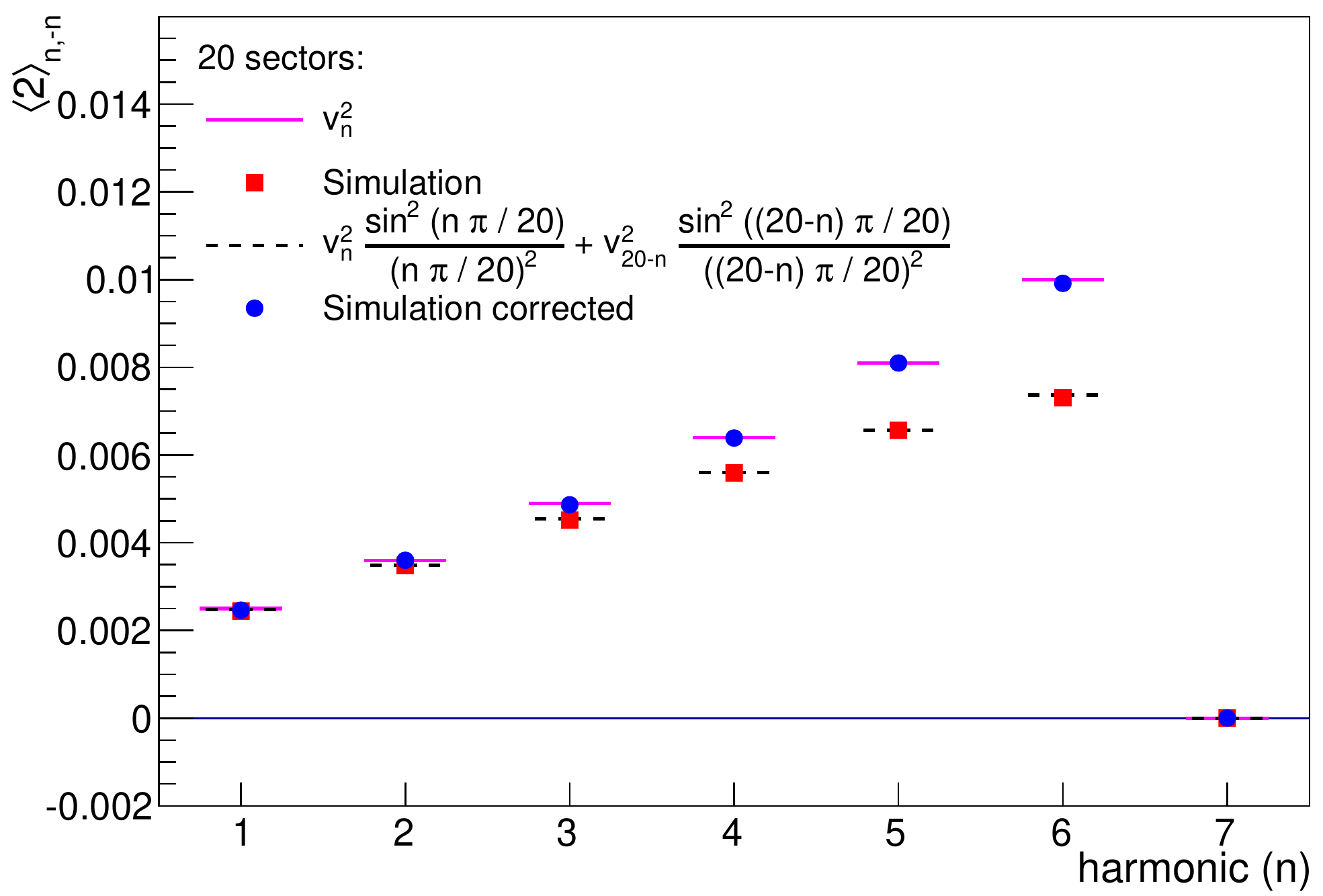}
\end{minipage}
\caption{The measured values of $\langle 2 \rangle_{n,-n}$ when the first 6 harmonics exist for detectors with 8, 12,
and 20 sectors are shown (red squares). The values of the input harmonics squared are represented by the magenta lines. The black dashed lines
show the expected measured value with that number of sectors. The blue circles show the result obtained when correcting
for the reduction to the measured value for the harmonic being measured.}
\label{fig:v2-6nsectors}
\end{figure}

If harmonics above $N/2$ are significant, Eq.~(\ref{eq:harmonic_expectation}) shows that finite segmentation will introduce
an interference from other harmonics (in fact, from an infinite number of harmonics). If one, for example, considers the case where the
first $N$ harmonics are non-zero, there will be a contribution from 2 terms in Eq.~(\ref{eq:harmonic_expectation}). As an example, we will once again consider the case
of the p.d.f in Eq.~(\ref{eq:pdf}) where the values of the first 6 harmonics are as in Eq.~(\ref{eq:inputValues}). In general if one considers the case where the
first $N$ harmonics are non-zero, then Eq.~(\ref{eq:harmonic_expectation}) produces the following relationship for $\langle 2 \rangle_{n,-n}$
when a factorizable p.d.f. (\ref{eq:factorization}) exists and one averages over many events:
\begin{equation}
E\left[\left<2\right>_{n,-n}\right] = v_{n}^{2} \frac{\sin^{2}\left(\frac{n \pi}{N}\right)}{\left(\frac{n \pi}{N}\right)^{2}} + v_{N-n}^{2} \frac{\sin^{2}\left(\frac{\left(n-N\right) \pi}{N}\right)}{\left(\frac{\left(n-N\right) \pi}{N}\right)^{2}} = E\left[\left<2\right>_{N-n,n-N}\right]\,.
\label{eq:2partcorrmultharm}
\end{equation}
The harmonic $v_{N-n}$, therefore, contaminates the measurement of $v_{n}$, although the harmonic below $N/2$ is dominant
(i.e. is suppressed less). For low segmentation, this can cause significant interference from other harmonics. If one tries to compute $v_2$
with 8 sectors, for instance, then $v_{N-n}$ corresponds to $v_6$ which contaminates the $v_2$ calculation. Eq.~(\ref{eq:2partcorrmultharm})
also explains the origin of the `blip' at $n=N/2$ in Fig.~\ref{fig:v2vsnsectors}, where both
terms are proportional to the same harmonic. Figure~\ref{fig:v2-6nsectors} shows this interference for a detector with 8, 12, and 20
azimuthal sectors. For 8 sectors, only $\langle 2 \rangle_{4,-4}$ can be corrected for. All other harmonics are contaminated
and it is easily seen that $\langle 2 \rangle_{1,-1} = \langle 2 \rangle_{7,-7}$, $\langle 2 \rangle_{2,-2} = \langle 2 \rangle_{6,-6}$,
and $\langle 2 \rangle_{3,-3} = \langle 2 \rangle_{5,-5}$ for the measured values. These cases cannot be corrected for. However, unless the
high harmonics are larger than the lower ones, the measured value will be closest to the lowest harmonic (in the example $v_{1}$ could
be corrected exactly only because $v_{7}$ was 0 in our toy MC study (\ref{eq:inputValues})). For 12 sectors, all of the existing harmonics can be calculated (and corrected for
finite granularity). However, $v_{7}^{2}$ is still calculated incorrectly because it is actually measuring $v_{1}^{2}$. For the case
of 20 sectors, all contaminations have disappeared and one can accurately determine the harmonics. In general, one can only measure
up to $v_{N/2}$ and one should have a reasonable estimate of the size of the other harmonics to determine if the contamination from harmonics with $n>N/2$ will be significant.

%


\section{Systematic bias due to particle selection criteria}
\label{s:Systematic bias due to particle selection criteria}

In this section we discuss our final topic which concerns the results based on multi-particle correlation techniques. In particular, we point out the existence of a systematic bias in traditional differential flow analyses with two- and multi-particle cumulants, which stems solely from the selection criteria applied on reference particles (RP) and on particles of interest (POI), and which is present also in the ideal case when all nonflow correlations are absent.
We need two separate groups of particles, RPs and POIs, in the traditional differential flow analyses to get a statistically stable result in the cases where there exists a small number of POIs in a narrow differential bin of interest.
The direct evaluation of the multi-particle correlators in Eq.~(\ref{eq:mpCorrelation}) using only POIs would result in statistically unstable results. To circumvent this, Borghini {\it et al} proposed in~\cite{Borghini:2000sa,Borghini:2001vi} to use RPs for all particles {\it except the first} in two- and multi-particle correlators, where RPs are selected from some large statistical sample of particles in an event (e.g. from all charged particles). Any dependence of the differential flow of POIs on RPs would  then be eliminated by separately evaluating multi-particle correlators by using only RPs and then by explicitly dividing out their corresponding contribution to differential multi-particle correlators, in which only the first particle was restricted to be a POI. For a detailed description of traditional differential flow analyses we refer the reader to~\cite{Borghini:2000sa, Borghini:2001vi}; now we quantify the systematic bias which stems solely from the applied selection criteria on RPs and POIs.
   
Usually it is said that collective anisotropic flow measured with $QC\{2\}$ is enhanced by flow fluctuations and $QC\{4\}$ is suppressed by flow fluctuations. When only using reference flow it is also easily shown that (for a detailed derivation, see Appendix~A in~\cite{Bilandzic:2012wva}):
\begin{eqnarray}
v\{2\} = \langle v \rangle + \frac{1}{2} \frac{\sigma_v^2}{\langle v \rangle} 
\label{eq:reference QC2}\,,\\
v\{4\} = \langle v \rangle - \frac{1}{2} \frac{\sigma_v^2}{\langle v \rangle}
\label{eq:reference QC4}\,,
\end{eqnarray}
where $\langle v \rangle$ is the mean value of the flow moment of interest, and $\sigma_v^2$ the variance of that flow moment. However, in the more generally applied case, where the reference flow is used to obtain a differential flow, the situation becomes more complicated. 

\subsection{$v'\{2\}$}
The differential 2-particle cumulant estimate, $v'\{2\}$, is obtained as~\cite{Borghini:2001vi}:
\begin{equation}
v'\{2\} = \frac{\langle v'v \rangle}{\sqrt{\langle v^2 \rangle}}\,.
\end{equation}
Using $\langle v'v \rangle = \langle v'\rangle \langle v \rangle + \rho \, \sigma_{v'} \sigma_v$, where $\rho$ is the correlation coefficient between the reference flow and the differential flow and is defined in the range $[-1,1]$, where specifically $\rho=1$ when $v$ and $v'$ are perfectly correlated, $\rho=0$ when they are uncorrelated, and $\rho=-1$ when they are anticorrelated, one can find the following relation assuming $\sigma_v ^2 / \langle v \rangle ^2 \ll 1$:
\begin{equation}
v'\{2\} \approx \langle v'\rangle \left(1+\rho\frac{\sigma_{v'}\sigma_{v}}{\langle v'\rangle \langle v \rangle}-\frac{1}{2}\frac{\sigma_v^2}{\langle v\rangle ^2}\right)\,, 
\label{eq:flow fluc 2-p}
\end{equation}
from which it is seen that $v'\{2\}$ can be \emph{suppressed} by flow fluctuations.\\

\subsection{$v'\{4\}$}
The differential 4-particle cumulant estimate, $v'\{4\}$, is obtained as~\cite{Borghini:2001vi}:
\begin{equation}
v'\{4\} = \frac{-\langle v' v^3 \rangle + 2\langle v'v\rangle \langle v^2 \rangle}{\left (v\{4\} \right)^3}\,.
\end{equation}
Using Eq.~(\ref{eq:reference QC4}) and $Var\left[f(x)\right]\approx \left( f'\left(E[x]\right)\right) ^2Var[x]$ one can obtain:
\begin{equation}
v'\{4\} \approx \langle v'\rangle \left(1-\rho\frac{\sigma_{v'}\sigma_{v}}{\langle v'\rangle \langle v \rangle}+\frac{1}{2}\frac{\sigma_v^2}{\langle v\rangle ^2}\right)\,. 
\label{eq:flow fluc 4-p}
 \end{equation}
This is very similar to Eq.~(\ref{eq:flow fluc 2-p}). Once again it is clear that the bias to the differential flow may not be the same as for the reference flow, an enhancement \emph{or} a suppression is possible. Three cases are explored in more detail below, while details of the calculations are provided in Appendix~\ref{s:Appendix to fluctuations}.\\

\subsection{Specific cases}
\noindent\textbf{$v'$ and $v$ are perfectly correlated ($\rho=1$) and $\sigma_{v'}/v' = \sigma_v/v$.} For this case, RPs and POIs can have a full overlap, but it is not required. Eq.~(\ref{eq:flow fluc 2-p}) can be written as:
\begin{equation}
v'\{2\} \approx \langle v'\rangle \left(1+\frac{\sigma_{v'}^2}{\langle v'\rangle^2}-\frac{1}{2}\frac{\sigma_{v'}^2}{\langle v'\rangle ^2}\right) =\langle v'\rangle \left(1+ \frac{1}{2}\frac{\sigma_{v'}^2}{\langle v' \rangle ^2} \right)\,.
\label{eq:case 1}
\end{equation}
This case reduces to the regular case where $QC\{2\}$ is systematically {\it enhanced} by flow fluctuations, just as for the reference flow. Since $v'\{4\}$ simply has opposite signs on the fluctuation terms, it follows that in this case it is \emph{suppressed}, once again the same as the reference flow.\\

\noindent\textbf{$v'$ and $v$ are uncorrelated ($\rho=0$).} In reality this covers a case where the RPs or POIs are chosen from a two groups of particles that do not overlap and do not contain the same underlying correlations. 
For this case $\rho = 0$, so Eq.~(\ref{eq:flow fluc 2-p}) trivially turns into:
\begin{equation}
v'\{2\} \approx \langle v'\rangle \left(1 - \frac{1}{2}\frac{\sigma_v^2}{\langle v\rangle ^2} \right)\,.
\label{eq:case 2}
\end{equation}
This means the differential 2-particle cumulant is systematically {\it suppressed} by the flow fluctuations in the reference flow, and that the 4-particle differential cumulant is systematically \emph{enhanced}. Fluctuations from the POIs do not play any role.\\

\noindent\textbf{$v'$ and $v$ are correlated, but the relative fluctuations are different.} Once again the RPs and POIs may have a full overlap, but it is not required. In this case it is assumed that $\rho\approx1$, leading to:
\begin{equation}
v'\{2\} \approx \langle v'\rangle \left(1+\frac{\sigma_{v'}\sigma_{v}}{\langle v'\rangle \langle v \rangle}-\frac{1}{2}\frac{\sigma_v^2}{\langle v\rangle ^2}\right)\,, 
\label{eq:case 3}
\end{equation}
and the observed bias for the 2-particle (4-particle) differential cumulant is an \emph{enhancement} (\emph{suppression}) as long as $2\left(\frac{\sigma_{v'}}{\langle v' \rangle}\right) > \left(\frac{\sigma_v}{\langle v \rangle}\right)$. In general the bias observed in the differential flow is influenced by the fluctuations in the reference flow.\\
\begin{figure}
\centering
\includegraphics[width=0.5\textwidth]{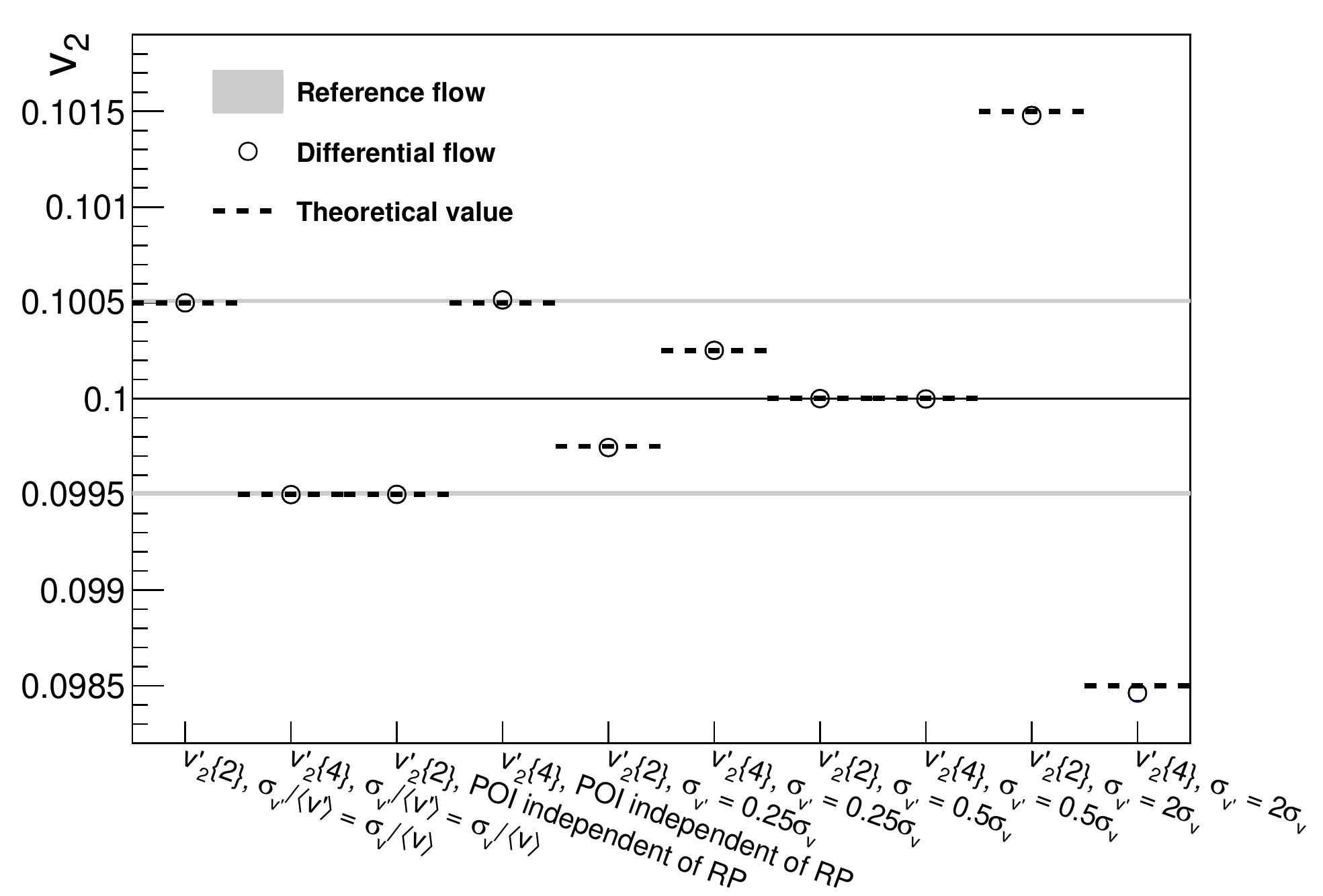}
\caption{$10^6$ events with 10000 RPs and 1000 POIs. Input flow is $v_2 = 0.1$, reference flow fluctuations have $\sigma_{v_2} = 0.01$. Depending on the choice of particles for differential flow and the differential flow fluctuations, it is possible to get very different biases to the 2- and 4-particle cumulants.}
\label{fig:flow fluc}
\end{figure}

To illustrate the different cases a simulation of $10^6$ events with 10000 RPs and 1000 POIs has been made. The results are shown in Fig.~\ref{fig:flow fluc} with input values, $v_2 = 0.1$ and $\sigma_{v_2} = 0.01$. In the figure Gaussian fluctuations are assumed, but other fluctuations, e.g., uniform fluctuations, would yield similar results. The shaded bands indicate the reference flow of $v\{2\}$ and $v\{4\}$, calculated with Eqs.~(\ref{eq:reference QC2}) and (\ref{eq:reference QC4}) respectively, showing the usual enhancement or suppression. The first two points are from a simulation illustrating the first case above, where the POIs and RPs are perfectly correlated and share the same relative fluctuations and have a full overlap. The dotted lines are calculated using Eq.~(\ref{eq:case 1}) for $QC\{2\}$ and the corresponding equation for $QC\{4\}$. For the next two points the POIs and RPs are chosen with independent fluctuations and no overlap. In this case $v'\{2\}$ and $v'\{4\}$ are swapped, as expected from Eq.~(\ref{eq:case 2}). The last points show cases where the relative fluctuations in the POIs differ from those in the RPs, this can cause the usual enhancement and suppression to be larger, swapped or even be removed completely, depending on how the relative fluctuations are chosen. In the example simulations shown here, RPs and POIs do not overlap. For the case where $\sigma_{v'}=0.25\sigma_v$ Eq.~(\ref{eq:case 3}) yields:
\begin{equation}
v'\{\substack{2 \\ 4}\} = \langle v' \rangle \left (1 \mp \frac{1}{4} \frac{\sigma_{v}^2}{\langle v \rangle^2}\right )\,.
\end{equation}
For $\sigma_{v'}=0.5\sigma_v$ Eq.~(\ref{eq:case 3}) yields:
\begin{equation}
v'\{\substack{2 \\ 4}\} = \langle v' \rangle\,, 
\end{equation}
and finally for $\sigma_{v'}=2\sigma_v$:
\begin{equation}
v'\{\substack{2 \\ 4}\} = \langle v' \rangle \left (1 \pm \frac{3}{2} \frac{\sigma_{v}^2}{\langle v \rangle^2}\right )\,.
\end{equation}
It is tempting to use Eqs.~(\ref{eq:reference QC2}) and (\ref{eq:reference QC4}) to estimate the magnitude of the flow fluctuations. However, when doing differential flow analysis with cumulants it is clear from Eqs.~(\ref{eq:flow fluc 2-p}) and (\ref{eq:flow fluc 4-p}) that it may not be feasible. In fact, any analysis using differential flow should be very careful to describe the choice of RPs and POIs in great detail, such that comparison between different experiments and theories is not biased by mixing two or more of the cases shown in Fig.~\ref{fig:flow fluc} and described above.


\section{Summary}
\label{s:summary}

We have presented the new generic framework within which all multi-particle azimuthal correlations can be evaluated analytically, with a fast single pass over the particles, free from autocorrelations by definition, and corrected for systematic biases due the various detector inefficiencies. For higher order correlators the direct implementation of analytic solutions is not feasible due to their size; this issue was resolved with the development of new recursive algorithms. We have proposed new multi-particle observables to be used in anisotropic flow analyses (standard candles) which can be measured for the first time within our generic framework. The systematic biases due to finite granularity of detector on multi-particle correlators have been quantified. We have pointed out the existence of a systematic bias characteristic for traditional differential flow analyses when all particles are divided into two groups of reference particles (RP) and particles of interest (POI), which originates solely from the selection criteria for RPs and POIs, and which is present also in the ideal case when all nonflow correlations are absent. Finally, we have straightforwardly generalized our generic framework to the case of differential multi-particle correlators.   

\acknowledgments{
We thank Jens J\o rgen Gaardh\o je for carefully reading the paper and providing valuable feedback for its improvement. We thank Jean-Yves Ollitrault and Sergei Voloshin for important suggestions in the historical account on multi-particle correlation techniques presented in the Introduction. We appreciate help and cheerful discussions from Marek Chojnacki in the derivation of recursive algorithms. We thank Jiangyong Jia for valuable feedback on the studies of correlations of event-by-event fluctuations of various harmonics. We thank the Danish Council for Independent Research, Natural Sciences and the Danish National Research Foundation (Danmarks Grundforskningsfond) for support. This work is also supported by FOM and NWO of the Netherlands.}

\appendix

\section{Recursive algorithm}
\label{s:Algorithm}

\noindent
As mentioned in Section~\ref{ss:Algorithm}, we provide implementations
\cite{CHC} for calculating generic multi-particle correlators defined in Eq.~(\ref{eq:mpCorrelation}) for:
\begin{description}
\item[\emph{Fully expanded}] expressions for
  $\Num{m}{n_{1},\ldots,n_{m}}$ (see (\ref{eq:num}) and (\ref{eq:2pCorrelation}-\ref{eq:4pCorrelation})) for $m=2,\ldots,8$;

\item[\emph{Recurrence}] expression
  $\Num{m}{n_{1},\ldots,n_{m}}^{\prime}$ (see \eqref{eq:algo1}) for any  $m$;
\item[\emph{Recursive}] expression
  $\Num{m}{n_{1},\ldots,n_{m}}^{\prime\prime}$ (see \eqref{eq:algo2}) for
  any $m$.
\end{description}
The largest feasible $m$ for the two latter methods above is of course limited
by computing time, resources and machine precision.  However, there
is no inherent limitations on $m$ in the implementations.

The implementation is done in plain callable C++ with no external
dependencies.  It can be integrated into any existing framework,
including ROOT~\cite{ROOT} based ones, by simple inclusion of the appropriate
headers.  Examples of standalone and ROOT applications are provided
in the code.  
The code itself is further heavily documented at \cite{CHC}. 

\begin{figure}[htbp]
  \centering
  \includegraphics[keepaspectratio,width=.5\linewidth]{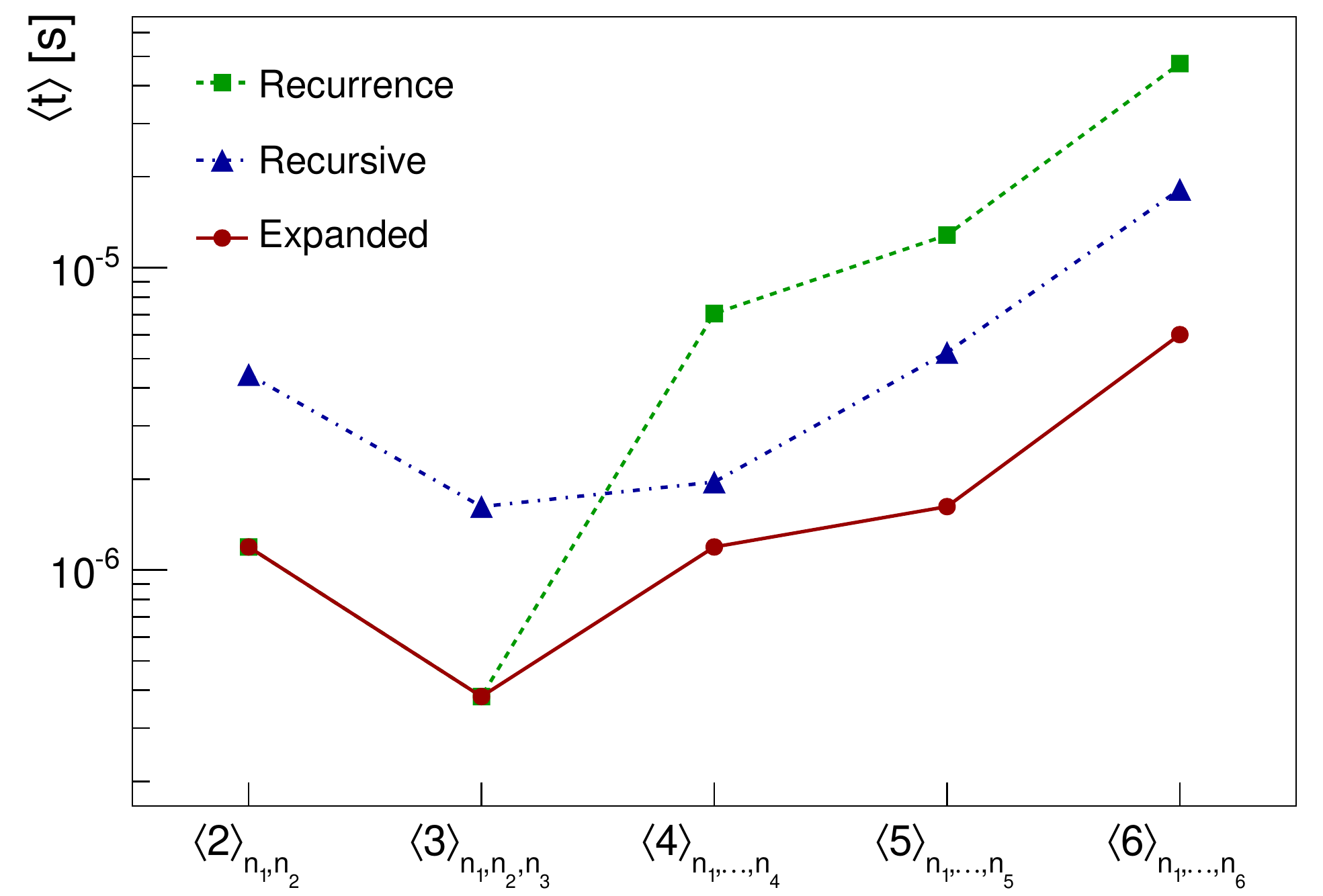}
  \caption[Average computation time (in seconds) of multi-particle correlators
  for the three methods as a function of correlator order]{Average
    computation time (in seconds) of multi-particle correlators for the
    three methods as a function of correlator order $m$: \emph{Fully
      expanded} are red circles, \emph{recurrence} are green squares,
    and \emph{recursive} are blue triangles.}
  \label{fig:app:algorithmsTiming}
\end{figure}

The choice of method, using either \emph{expanded}, \emph{recurrence},
or \emph{recursive} expression, is left to the user.  However, it
should be noted, that using the truly general \emph{recurrence},
or \emph{recursive} expressions does incur a performance penalty,
as can be seen from Fig.~\ref{fig:app:algorithmsTiming}. 

\section{Finite granularity}
\label{s:Appendix_to_finite_granularity}

In this appendix, the equations used in Section~\ref{s:Detectors with finite granularity} to evaluate the effects of finite granularity are derived. We start by defining a
detector with $N$ equal size adjacent azimuthal sectors with sectors being labeled by an integer $i$ where $0 \leq i \leq N-1$.
Furthermore the low edge of the first sector is shifted from 0 by $\varphi_{\Delta}$. The edges of sector $i$ are then defined by:
\begin{equation}
\varphi_{L_{i}} = i\frac{2\pi}{N}+\varphi_\Delta,\qquad \varphi_{H_{i}} = (i+1)\frac{2\pi}{N}+\varphi_\Delta \,.
\end{equation}
The p.d.f. for any particle is taken to be:
\begin{equation}
\frac{dP}{d\varphi} = \frac{1}{2 \pi} \left[1+\sum_{n=1}^{\infty}2v_n\cos\left(n\left(\varphi-\Psi_n\right)\right)\right]\,.
\end{equation}
The probability of a particle going into sector $i$ is then found by integrating over the limits of the sector:
\begin{eqnarray}
P_i & = \int_{\varphi_{L_{i}}}^{\varphi_{H_{i}}} \frac{dP}{d\varphi} d\varphi & = \frac{1}{2 \pi} \left[ \int_{\varphi_{L_{i}}}^{\varphi_{H_{i}}} d\varphi + \sum_{n=1}^{\infty} 2v_n \int_{\varphi_{L_{i}}}^{\varphi_{H_{i}}} \cos\left(n\left(\varphi-\Psi_n\right)\right) d\varphi \right] \notag\\
& & = \frac{1}{2 \pi} \left[ \frac{2 \pi}{N} + \sum_{n=1}^{\infty} 2v_n \frac{\sin \left(n\left(\varphi_{L_{i}} - \Psi_n\right)\right) - \sin \left(n\left(\varphi_{H_{i}} - \Psi_n\right)\right)}{n} \right] \notag\\
& & = \frac{1}{2 \pi} \left[ \frac{2 \pi}{N} + \sum_{n=1}^{\infty} 2v_n \frac{2 \sin \left(n\frac{\left(\varphi_{H_{i}} - \varphi_{L_{i}}\right)}{2}\right) \cos \left(n\left(\frac{\varphi_{H_{i}}+\varphi_{L_{i}}}{2} - \Psi_n\right)\right)}{n} \right] \notag\\
& & = \frac{1}{N} \left[ 1 + \sum_{n=1}^{\infty} 2v_n \frac{\sin \left(n\frac{\left(\varphi_{H_{i}} - \varphi_{L_{i}}\right)}{2}\right)}{\frac{n \pi}{N}} \cos \left(n\left(\frac{\varphi_{H_{i}}+\varphi_{L_{i}}}{2} - \Psi_n\right)\right) \right] \notag\\
& & = \frac{1}{N} \left[ 1 + \sum_{n=1}^{\infty} 2v_n \frac{\sin \frac{n \pi}{N}}{\frac{n \pi}{N}} \cos \left(n\left(\left(i+\frac{1}{2}\right)\frac{2 \pi}{N} + \varphi_\Delta - \Psi_n\right)\right) \right]\,.
\end{eqnarray}
The expected value of $e^{im\varphi}$ must then be evaluated as follows:
\begin{eqnarray}
E\left[ e^{im\varphi} \right] & = & \sum\limits_{j=0}^{N-1} e^{im\left[\left(j+\frac{1}{2}\right)\frac{2\pi}{N}+\varphi_\Delta\right]} P_j \notag\\
& = & \frac{1}{N} \sum_{j=0}^{N-1} e^{im\left[(j+\frac{1}{2})\frac{2\pi}{N}+\varphi_{\Delta}\right]} \notag\\
& &{}+\frac{1}{N}\sum_{n=1}^{\infty}v_n\frac{\sin\frac{n\pi}{N}}{\frac{n\pi}{N}}
\left[ e^{-in\Psi_{n}} \sum_{j=0}^{N-1}e^{i(m+n)\left[(j+\frac{1}{2})\frac{2\pi}{N}+\varphi_{\Delta}\right]}
+ e^{in\Psi_{n}}\sum_{j=0}^{N-1}e^{i(m-n)\left[(j+\frac{1}{2})\frac{2\pi}{N}+\varphi_{\Delta}\right]}\right]\,.
\label{eq:expectexp}
\end{eqnarray}
%

%
%
Eq.~(\ref{eq:expectexp}) has terms of the form $\sum_{j=0}^{N-1} e^{ik\left[(j+\frac{1}{2})\frac{2\pi}{N}+\varphi_{\Delta}\right]}$,
where $k$ is an integer, that must be evaluated. We can first evaluate the following:
\begin{eqnarray}
\sum_{j=0}^{N-1}e^{i(j+\frac{1}{2})\frac{2\pi k}{N}} & = &
e^{i\frac{\pi k}{N}} + e^{3i\frac{\pi k}{N}} + \ldots + e^{(2N-3)i\frac{\pi k}{N}}+e^{(2N-1)i\frac{\pi k}{N}} \notag\\
& = & e^{i\frac{\pi k}{N}} \left\{ 1 + e^{2i\frac{\pi k}{N}} + \ldots + e^{2(N-2)i\frac{\pi k}{N}}+e^{2(N-1)i\frac{\pi k}{N}} \right\}\,.
\label{eq:exp_sum} 
\end{eqnarray}
If $k/N \in \mathbb{Z}$, where $\mathbb{Z}$ is the set of all integers, then:
\begin{eqnarray}
\sum_{j=0}^{N-1}e^{i(j+\frac{1}{2})\frac{2\pi k}{N}} & \stackrel{\frac{k}{N}\in\mathbb{Z}}{=} & (-1)^{\frac{k}{N}} \left\{ 1 + 1 + \ldots + 1 + 1 \right\} \notag\\
& \stackrel{\frac{k}{N}\in\mathbb{Z}}{=} & N(-1)^{\frac{k}{N}}\,.
\label{eq:exp_sum_multiple}
\end{eqnarray}
If $k/N \notin \mathbb{Z}$ then the sum can be evaluated as follows:
\begin{equation*}
\underbrace{\left(1-e^{2\pi i\frac{k}{N}}\right)}_{\text{not 0 if $\frac{k}{N} \notin \mathbb{Z}$}}\times
\sum_{j=0}^{N-1}e^{i(j+\frac{1}{2})\frac{2\pi k}{N}} \stackrel{\frac{k}{N}\notin\mathbb{Z}}{=} e^{i\frac{\pi k}{N}}\underbrace{\left(1-e^{2\pi ik}\right)}_{\text{0 since $k \in \mathbb{Z}$}}\,,
\end{equation*}
which yields:
\begin{equation}
\sum_{j=0}^{N-1}e^{i(j+\frac{1}{2})\frac{2\pi k}{N}} \stackrel{\frac{k}{N}\notin\mathbb{Z}}{=} 0\,.
\label{eq:exp_sum_not_multiple}
\end{equation}
Therefore the following is true:
\begin{equation}
\sum_{j=0}^{N-1}e^{ik(j+\frac{1}{2})\frac{2\pi}{N}} = \left\{\begin{tabular}{cl}$N(-1)^{\frac{k}{N}}$ & for $\frac{k}{N} \in \mathbb{Z}$\\0 & for $\frac{k}{N} \notin \mathbb{Z}$\end{tabular}\right.
\end{equation}
from which it follows:
\begin{equation} \sum_{j=0}^{N-1}e^{ik\left[(j+\frac{1}{2})\frac{2\pi}{N}+\varphi_\Delta\right]} = \left\{\begin{tabular}{cl}$N(-1)^{\frac{k}{N}}e^{ik\varphi_\Delta}$ & for $\frac{k}{N} \in \mathbb{Z}$\\0 & for $\frac{k}{N} \notin \mathbb{Z}$\end{tabular}\right.
\label{eq:exp_sum2}
\end{equation}
If we then define the following function:
\begin{equation}
\alpha(a,b) \equiv \left\{\begin{tabular}{cl}$(-1)^{\frac{a}{b}}$ & for $\frac{a}{b} \in \mathbb{Z}$\\0 & for $\frac{a}{b} \notin \mathbb{Z}$\end{tabular}\right.
\end{equation}
then Eq.~(\ref{eq:expectexp}) becomes:
\begin{equation}
E\left[ e^{im\varphi} \right] = e^{im\varphi_{\Delta}} \alpha(m,N)
+\sum_{n=1}^{\infty}v_n\frac{\sin\frac{n\pi}{N}}{\frac{n\pi}{N}}
\left[ e^{-in\Psi_{n}} e^{i\left(m+n\right)\varphi_{\Delta}} \alpha(m+n,N)
+ e^{in\Psi_{n}} e^{i\left(m-n\right)\varphi_{\Delta}} \alpha(m-n,N) \right]\,.
\label{eq:expectexp2}
\end{equation}

%

The terms with $\alpha(m+n,N)$ will have $m+n = jN$ where $j$ is an integer. Values of $n$ which produce non-zero contributions must
have $n = jN-m$. Since $n>0$, this means that $jN-m>0$ and, therefore, $j>m/N$. The following relation is then true:
\begin{eqnarray}
&& \sum_{n=1}^{\infty} v_n\frac{\sin\frac{n\pi}{N}}{\frac{n\pi}{N}}e^{-in\Psi_{n}} e^{i\left(m+n\right)\varphi_{\Delta}} \alpha(m+n,N)\notag\\
&& \hphantom{\sum_{n=1}^{\infty}} = \sum_{\begin{subarray}{c}j =-\infty\\j > \frac{m}{N}\end{subarray}}^{\infty} v_{jN-m}\frac{\sin\frac{\left(jN-m\right)\pi}{N}}{\frac{\left(jN-m\right)\pi}{N}} e^{-i\left(jN-m\right)\Psi_{jN-m}} e^{ijN\varphi_{\Delta}} (-1)^{j}\,.
\end{eqnarray}
The same argument can be made for the $\alpha(m-n,N)$ terms where $n = m-jN$ and $j<m/N$ giving the following relation:
\begin{eqnarray}
&& \sum_{n=1}^{\infty} v_n\frac{\sin\frac{n\pi}{N}}{\frac{n\pi}{N}}e^{in\Psi_{n}} e^{i\left(m-n\right)\varphi_{\Delta}} \alpha(m-n,N)\notag\\
&& \hphantom{\sum_{n=1}^{\infty}} = \sum_{\begin{subarray}{c}j =-\infty\\j < \frac{m}{N}\end{subarray}}^{\infty} v_{m-jN}\frac{\sin\frac{\left(m-jN\right)\pi}{N}}{\frac{\left(m-jN\right)\pi}{N}} e^{i\left(m-jN\right)\Psi_{m-jN}} e^{ijN\varphi_{\Delta}} (-1)^{j}\,.
\end{eqnarray}
The two sets of terms can be combined as follows:
\begin{eqnarray}
&& \sum_{n=1}^{\infty} v_n\frac{\sin\frac{n\pi}{N}}{\frac{n\pi}{N}} \left[ e^{-in\Psi_{n}} e^{i\left(m+n\right)\varphi_{\Delta}} \alpha(m+n,N) + e^{in\Psi_{n}} e^{i\left(m-n\right)\varphi_{\Delta}} \alpha(m-n,N) \right]\notag\\
&& \hphantom{\sum_{n=1}^{\infty}} = \sum_{\begin{subarray}{c}j =-\infty\\j \neq \frac{m}{N}\end{subarray}}^{\infty} v_{\left|jN-m\right|}\frac{\sin \left(j-\frac{m}{N}\right)\pi}{\left(j-\frac{m}{N}\right)\pi} (-1)^{j} e^{-i\left(jN-m\right)\Psi_{\left|jN-m\right|}} e^{ijN\varphi_{\Delta}}\,.
\end{eqnarray}
With this relation, Eq.~(\ref{eq:expectexp2}) becomes:
\begin{equation}
E\left[e^{im\varphi}\right] =
 e^{im\varphi_\Delta}\alpha(m,N)+\sum_{\begin{subarray}{c}j =-\infty\\j \neq \frac{m}{N}\end{subarray}}^{\infty} v_{\left|jN-m\right|}\frac{\sin \left(j-\frac{m}{N}\right)\pi}{\left(j-\frac{m}{N}\right)\pi} (-1)^j e^{-i\left(jN-m\right)\Psi_{\left|jN-m\right|}} e^{ijN\varphi_{\Delta}}\,.
\label{eq:expectexp3}
\end{equation}
If $m$ is a multiple of $N$, every term in the second part of Eq.~(\ref{eq:expectexp3}) is 0 either because the term
is excluded ($j \neq m/N$) or because $\sin (j-m/N) \pi = 0$. The first term, however, is 0 if $m$ is not a multiple of $N$.
The two sets of terms, therefore, contribute to mutually exclusive sets of values of $m$. Eq.~(\ref{eq:expectexp3})
can then be rewritten as:
\begin{equation}
E\left[e^{im\varphi}\right] = \left\{
\begin{tabular}{ll}
$(-1)^{\frac{m}{N}} e^{im\varphi_\Delta}$ & for $\frac{m}{N} \in \mathbb{Z}$\\
$\sum\limits_{j=-\infty}^{\infty} v_{\left|jN-m\right|}\frac{\sin(j-\frac{m}{N})\pi}{(j-\frac{m}{N})\pi}(-1)^j e^{-i\left\{(jN-m)\Psi_{\left|jN-m\right|}-jN\varphi_{\Delta}\right\}}$ & for $\frac{m}{N} \notin \mathbb{Z}$\,.
\end{tabular}
\right. 
\label{eq:expectecp4}
\end{equation}
The asymptotic behavior of $E\left[e^{im\varphi}\right]$ agrees with what is expected. If $m=0$, $m$ is always a
multiple of $N$ and one should use the equation for $\frac{m}{N} \in \mathbb{Z}$ with $m=0$ which gives 1. If
$m \neq 0$, any fixed value of $m$ will not be a multiple of $N$ as $N \rightarrow \infty$ and one should use
the equation for $\frac{m}{N} \notin \mathbb{Z}$. As $N \rightarrow \infty$, all other terms, except for the
$j=0$ term, become 0 because $\sin (j-m/N) \pi \rightarrow 0$. The $j=0$ term has
$\sin\left(-\frac{m}{N}\pi\right)/\left(-\frac{m}{N}\pi\right) \rightarrow 1$ as $N \rightarrow \infty$ leaving
a value of $v_m e^{im\Psi_{m}}$.


\section{Systematic bias due to particle selection criteria}
\label{s:Appendix to fluctuations}

As mentioned in Section~\ref{s:Systematic bias due to particle selection criteria} for reference flow:
\begin{eqnarray}
v\{2\} = \langle v \rangle + \frac{1}{2} \frac{\sigma_v^2}{\langle v \rangle}\,, 
\label{eq:reference QC2 app}
\\
v\{4\} = \langle v \rangle - \frac{1}{2} \frac{\sigma_v^2}{\langle v \rangle}\,,
\label{eq:reference QC4 app}
\end{eqnarray}
where $\langle v \rangle$ is the mean value of the flow moment of interest and $\sigma_v^2$ is the variance of that flow moment. This can be obtained by assuming $\sigma_v ^2 / \langle v \rangle ^2 \ll 1$ and using~\cite{Bilandzic:2012wva}:
\begin{equation}
\langle f(x) \rangle \equiv E[f(x)] \approx f(\mu_x) + \frac{\sigma_x^2}{2}f''(\mu_x)\,,
\label{eq:expectation value app}
\end{equation}
where $E[x]$ is the expectation value of a random variable $x$, $f(x)$ is any function, $\mu_x$ is the mean of $x$, and $\sigma_x$ is the standard deviation of $x$. Below the same calculations are done for the 2- and 4-particle differential cumulants. 

\subsection{$v'\{2\}$}
The differential 2-particle cumulant estimate, $v'\{2\}$, is obtained by~\cite{Borghini:2001vi}:
\begin{equation}
v'\{2\} = \frac{\langle v'v \rangle}{\sqrt{\langle v^2 \rangle}}\,,
\end{equation}
where $v$ is the flow moment of the reference particles (RPs) and $v'$ is the differential flow moment of the particles of interest (POIs). Inserting Eq.~(\ref{eq:reference QC2 app}) for $\sqrt{\left<v^2\right>}$ and again assuming $\sigma_v ^2 / \langle v \rangle ^2 \ll 1$ yields:
\begin{equation}
v'\{2\} \approx \frac{\langle v'v \rangle}{\langle v\rangle }\left(1-\frac{1}{2}\frac{\sigma_v^2}{\langle v\rangle ^2}\right)\,. 
\end{equation}
The main issue is then to determine $\langle v'v\rangle $. In general:
\begin{equation}
\langle v'v \rangle = \langle v'\rangle \langle v \rangle + \rho \, \sigma_{v'} \sigma_v\,,
\end{equation}
where $\rho$ is the correlation coefficient between the reference flow and the differential flow and is defined in the range $[-1,1]$, where specifically $\rho=1$ in the case where $v$ and $v'$ are perfectly correlated, $\rho=0$ when they are uncorrelated and $\rho=-1$ when they are anticorrelated. $\sigma_{v'}$ is the standard deviation of the flow moment for POIs. This means:
\begin{equation}
v'\{2\} \approx \langle v'\rangle \left(1+\rho\frac{\sigma_{v'}\sigma_{v}}{\langle v'\rangle \langle v \rangle}-\frac{1}{2}\frac{\sigma_v^2}{\langle v\rangle ^2}\right)\,, 
\label{eq:flow fluc 2-p app}
\end{equation}
from which it is clearly seen that $v'\{2\}$ can be either \emph{suppressed} or \emph{enhanced} by flow fluctuations depending on the value of $\rho$.\\

\subsection{$v'\{4\}$}
The differential 4-particle cumulant estimate, $v'\{4\}$, is obtained by~\cite{Borghini:2001vi}:
\begin{equation}
v'\{4\} = \frac{-\langle v' v^3 \rangle + 2\langle v'v\rangle \langle v^2 \rangle}{\left (v\{4\} \right)^{3}}\,.
\end{equation}
Using Eq.~(\ref{eq:reference QC4 app}) this becomes:
\begin{equation}
v'\{4\} = \frac{-\langle v' v^3 \rangle + 2\langle v'v\rangle \langle v^2 \rangle}{\langle v \rangle ^3} \left(1+\frac{3}{2}\frac{\sigma_v^2}{\langle v \rangle ^2} \right)\,.
\label{eq:KGintervention_2}
\end{equation}
$-\langle v' v^3 \rangle + 2\langle v'v\rangle \langle v^2 \rangle$ must now be estimated. By using:
\begin{eqnarray}
Var\left[f(x)\right]&\equiv&E[f(x)^2]-E[f(x)]^2\nonumber\\
&\approx& \left( f'\left(\mu_x\right)\right) ^2Var[x]\,,
\label{eq:variance-cubed}
\end{eqnarray}
then 
\begin{eqnarray}
\langle v' v^3 \rangle &=& \langle v' \rangle \langle v^3 \rangle + \rho' \sigma_{v'}\sigma_{v^3} \\
 &\approx& \langle v' \rangle \left ( \langle v \rangle ^3 +3\sigma_v^2\langle v \rangle \right ) + \rho \sigma_{v'}3\langle v \rangle ^2\sigma_v \\ 
 &=& \langle v' \rangle \langle v \rangle ^3 + 3 \langle v' \rangle \langle v \rangle \sigma_v^2 + 3 \rho \langle v \rangle ^2 \sigma_{v'} \sigma_v\,,
\end{eqnarray}
where Eq.~(\ref{eq:expectation value app}) was also used for $\langle v^3 \rangle$. $\rho'$ is the correlation between $\sigma_{v'}$ and $\sigma_{v^3}$, applying the approximation in Eq.~(\ref{eq:variance-cubed}) to get to $\sigma_v$ yields the correlation between $\sigma_{v'}$ and $\sigma_v$, which is $\rho$ to first order. The next term to be estimated:
\begin{eqnarray}
2\langle v'v \rangle\langle v^2 \rangle &=& 2\left ( \langle v' \rangle \langle v \rangle + \rho\sigma_{v'}\sigma_v \right )\left (\sigma_v^2 + \langle v \rangle^2 \right ) \\
 &=& 2\langle v'\rangle \langle v \rangle \sigma_v^2 + 2 \langle v'\rangle\langle v \rangle ^3 + 2 \rho\langle v \rangle^2 \sigma_v \sigma_{v'} + 
2\rho \sigma_{v'}\sigma_v^3
\label{eq:KGintervention}
\end{eqnarray}
The last term in Eq.~(\ref{eq:KGintervention}) can be neglected. 
Inserting these results into Eq.~(\ref{eq:KGintervention_2}) it is seen that flow fluctuations bias $v'\{4\}$ in the following way:
\begin{eqnarray}
v'\{4\} &\approx& \frac{\langle v' \rangle\langle v\rangle^3 - \langle v' \rangle\langle v \rangle \sigma_v^2 -\rho\langle v\rangle^2\sigma_{v'}\sigma_v}{\langle v \rangle ^3} \left(1+\frac{3}{2}\frac{\sigma_v^2}{\langle v \rangle ^2} \right) \\
 &=& \langle v'\rangle \left ( 1 - \frac{\sigma_v^2}{\langle v \rangle ^2}-\rho\frac{\sigma_{v'}\sigma_{v}}{\langle v'\rangle \langle v \rangle}\right ) \left(1+\frac{3}{2}\frac{\sigma_v^2}{\langle v \rangle ^2} \right) \\
 &\approx& \langle v'\rangle \left(1-\rho\frac{\sigma_{v'}\sigma_{v}}{\langle v'\rangle \langle v \rangle}+\frac{1}{2}\frac{\sigma_v^2}{\langle v\rangle ^2}\right) \,,
 \end{eqnarray}
which once again can lead to either \emph{suppression} or \emph{enhancement} of flow fluctuations. In general one can write:
\begin{equation}
v'\{\substack{2\\4}\} \approx \langle v' \rangle \left (1 \pm \rho\frac{\sigma_{v'}\sigma_{v}}{\langle v'\rangle \langle v \rangle} \mp \frac{1}{2}\frac{\sigma_v^2}{\langle v\rangle ^2}\right) \,,
\end{equation}
showing that the bias to the 2- and 4-particle cumulants are similar but opposite. 


\section{Differential multi-particle correlators}
\label{s:Appendix to differential multi-particle correlators}

In this appendix we present generic equations for the differential (or reduced) correlators up to and including order four. All particles which are taken for the analysis are divided in each event into two groups: Reference Particles (RP) and Particles of Interest (POI), which in general can overlap. In each differential multi-particle correlator we specify the first particle to be POI, and all remaining particles to be RP. By adopting the original notation introduced by Borghini {\it et al}~\cite{Borghini:2001vi}, we label azimuthal angles of POIs with $\psi$, and azimuthal angles of RPs with $\varphi$. In practice, POIs will correspond to particles in a differential bin of interest in an event (e.g. particles in a narrow $p_{\rm T}$ bin, particles in a narrow $\eta$ bin, etc.), while RPs correspond to some large statistical sample of particles in an event (e.g. all charged particles). 
 
The average differential $m$-particle correlation in harmonics $n_1,n_2,\ldots,n_m$ is given by the following generic definition:
\begin{eqnarray}
\left<m'\right>_{\underline{n_1},n_2,\ldots,n_m}&\equiv&\left<e^{i(n_1\psi_{k_1}+n_2\varphi_{k_2}+\cdots+n_m\varphi_{k_m})}\right>\nonumber\\
&\equiv&\frac{\displaystyle\sum_{k_1}^{m_p}  \displaystyle\sum_{\begin{subarray}{c}k_2,\ldots,k_m=1\\k_1\neq k_2\neq \ldots\neq k_m\end{subarray}}^{M}w_{k_1}w_{k_2}\cdots w_{k_m}\,e^{i(n_1\psi_{k_1}+n_2\varphi_{k_2}+\cdots+n_m\varphi_{k_m})}}{\displaystyle\sum_{k_1}^{m_p}\displaystyle\sum_{\begin{subarray}{c}k_2,\ldots,k_m=1\\k_1\neq k_2\neq \ldots\neq k_m\end{subarray}}^{M} w_{k_1}w_{k_2}\cdots w_{k_m}}\,.
\label{eq:mpDifferentialCorrelation}
\end{eqnarray}
In the above definition $M$ is the number of RPs in an event, $m_p$ is number of POIs in a narrow differential bin in an event, $\varphi$ labels the azimuthal angles of RPs, $\psi$ labels the azimuthal angles of POIs, while $w$ labels particle weights. In general, we allow independent particle weights for RPs and POIs. All trivial effects from autocorrelations are removed by the constraint $k_1\neq k_2\neq \ldots\neq k_m$, which enforces all indices in all summands to be unique in definition (\ref{eq:mpDifferentialCorrelation}). The only harmonic which corresponds to POIs is underlined, in order to distinguish it from the all other harmonics which correspond to RPs. 

As in the case of reference multi-particle correlators studied in the main part of this paper, we first observe that the expressions in the numerator and the denominator of Eq.~(\ref{eq:mpDifferentialCorrelation}) are trivially related. Therefore we introduce the following shortcuts: 
\begin{eqnarray}
\mathrm{N}\left<m'\right>_{\underline{n_1},n_2,\ldots,n_m}&\equiv&
\displaystyle\sum_{k_1}^{m_p}\displaystyle\sum_{\begin{subarray}{c}k_2,\ldots,k_m=1\\k_1\neq k_2\neq \ldots\neq k_m\end{subarray}}^{M}w_{k_1}w_{k_2}\cdots w_{k_m}\,e^{i(n_1\psi_{k_1}+n_2\varphi_{k_2}+\cdots+n_m\varphi_{k_m})}\,,\label{eq:numDiff}\\
\mathrm{D}\left<m'\right>_{\underline{n_1},n_2,\ldots,n_m}&\equiv&\displaystyle\sum_{k_1}^{m_p}\displaystyle\sum_{\begin{subarray}{c}k_2,\ldots,k_m=1\\k_1\neq k_2\neq \ldots\neq k_m\end{subarray}}^{M}\!w_{k_1}w_{k_2}\cdots w_{k_m}\label{eq:denDiff}\\
&=&\mathrm{N}\left<m'\right>_{\underline{0},0,\ldots,0}\,.\label{eq:denNumDiff0000}
\end{eqnarray}
We will present our results for expressions (\ref{eq:numDiff}) and (\ref{eq:denDiff}) in terms of weighted $Q$-, $p$- and $q$-vectors, that we now define. The weighted $Q$-vector is a complex number defined by
\begin{equation}
Q_{n,l} \equiv \sum_{k=1}^{M}w_k^l\,e^{in\varphi_k} \,,
\label{eq:Qvector:App}
\end{equation}
and filled with all particles labeled as RPs in an event ($M$ in total).
The weighted $p$-vector is constructed out of all POIs ($m_p$ in
total) in a narrow differential bin of interest in an event:
\begin{eqnarray}
p_{n,l} &\equiv& \sum_{k=1}^{m_p} w_k^l\,e^{in\psi_k}\,.
\label{p-vectorDefinition}
\end{eqnarray}
Lastly, the weighted $q$-vector is constructed only from particles in a narrow differential bin of interest in an event which are labeled both as POIs and RPs ($m_q$ in total):
\begin{eqnarray}
q_{n,l} &\equiv& \sum_{k=1}^{m_q} w_k^l\,e^{in\psi_k}\,.
\label{q-vectorDefinition}
\end{eqnarray}
The $q$-vector was introduced in order to analytically remove all effects of autocorrelations in our final results. The indices $n$ and $l$ in definitions (\ref{eq:Qvector:App})-(\ref{q-vectorDefinition}) are determined from the original  
indices $n_1,n_2,\ldots,n_m$ in (\ref{eq:mpDifferentialCorrelation}), as will become clear shortly. In general, we will need $Q$-, $p$- and $q$-vectors evaluated for multiple values of indices $n$ and $l$, which will be determined by the precise nature of the differential multi-particle correlator in question. The key point, however, is that to obtain $Q$-, $p$- and $q$-vectors for, in principle, any number of different values of indices $n$ and $l$, a single pass over all particles still suffices.  

Given the above definitions, and by following the same strategy and notation as in the main part of the paper, we have obtained the following analytic results for differential 2-, 3- and 4-particle correlations:
\begin{eqnarray}
\mathrm{N}\left<2'\right>_{\underline{n_1},n_2}&=&p_{n_1,1} Q_{n_2,1}-q_{n_1+n_2,2}\,,\\
\mathrm{D}\left<2'\right>_{\underline{n_1},n_2}&=&\mathrm{N}\left<2'\right>_{\underline{0},0}\nonumber\\ 
&=&p_{0,1} Q_{0,1}-q_{0,2}\,. 
\label{eq:2pDiffCorrelation}
\end{eqnarray}
\begin{eqnarray}
\mathrm{N}\left<3'\right>_{\underline{n_1},n_2,n_3}&=&p_{n_1,1} Q_{n_2,1} Q_{n_3,1}-q_{n_1+n_2,2} Q_{n_3,1}-q_{n_1+n_3,2}Q_{n_2,1}\nonumber\\
&&{}-p_{n_1,1} Q_{n_2+n_3,2}+2 q_{n_1+n_2+n_3,3}\,,\\
\mathrm{D}\left<3'\right>_{\underline{n_1},n_2,n_3}&=&\mathrm{N}\left<3'\right>_{\underline{0},0,0}\nonumber\\
&=&p_{0,1} Q_{0,1}^2-p_{0,1} Q_{0,2}-2q_{0,2} Q_{0,1}+2 q_{0,3}\,.
\label{eq:3pDiffCorrelation}
\end{eqnarray}
\begin{eqnarray}
\mathrm{N}\left<4'\right>_{\underline{n_1},n_2,n_3,n_4}&=&p_{n_1,1} Q_{n_2,1} Q_{n_3,1} Q_{n_4,1}-q_{n_1+n_2,2} Q_{n_3,1} Q_{n_4,1}
-q_{n_1+n_3,2} Q_{n_2,1} Q_{n_4,1}\nonumber\\
&&{}-p_{n_1,1} Q_{n_2+n_3,2} Q_{n_4,1}+2 q_{n_1+n_2+n_3,3} Q_{n_4,1}-q_{n_1+n_4,2}Q_{n_2,1}
Q_{n_3,1} \nonumber\\
&&{}+q_{n_1+n_4,2}Q_{n_2+n_3,2} 
-p_{n_1,1} Q_{n_3,1} Q_{n_2+n_4,2}+q_{n_1+n_3,2} Q_{n_2+n_4,2}\nonumber\\
&&{}+2 q_{n_1+n_2+n_4,3} Q_{n_3,1} 
-p_{n_1,1} Q_{n_2,1} Q_{n_3+n_4,2}+q_{n_1+n_2,2}
Q_{n_3+n_4,2}\nonumber\\
&&{}+2 q_{n_1+n_3+n_4,3}Q_{n_2,1} +2 p_{n_1,1} Q_{n_2+n_3+n_4,3}-6 q_{n_1+n_2+n_3+n_4,4}\,,\\
\mathrm{D}\left<4'\right>_{\underline{n_1},n_2,n_3,n_4}&=&\mathrm{N}\left<4'\right>_{\underline{0},0,0,0}\nonumber\\
&=&
p_{0, 1}Q_{0, 1}^3 - 3 q_{0, 2} Q_{0, 1}^2 - 3 p_{0, 1}Q_{0, 1} Q_{0, 2}+ 3 q_{0, 2} Q_{0, 2} + 6 q_{0, 3}Q_{0, 1} \nonumber \\
&& {}+ 2 p_{0, 1} Q_{0, 3} - 
 6 q_{0, 4} \,.
\label{eq:4pDiffCorrelation}
\end{eqnarray}

The above relations are generic equations for differential multi-particle correlators, and they improve and generalize over the limited results presented in~\cite{Bilandzic:2010jr}, which were applicable only for the special case in which all harmonics $n_1,n_2,\ldots,n_m$ coincide. The further improvement consists of the fact that with these new results we allow for an independent weighting of POIs and RPs straight from the definition (see Eqs. (\ref{eq:Qvector:App}) and
(\ref{p-vectorDefinition})) which will have an obvious use case in experimental analyses when reconstruction efficiency for POIs and RPs differs. Finally, we have preserved the full generality when it comes to different possible outcomes of particle labeling; the results above are applicable for all three distinct cases of labeling, namely ``no overlap", ``partial overlap" and ``full overlap", between RPs and POIs. 


\end{document}